\documentclass[runningheads]{llncs}
\usepackage[T1]{fontenc}
\usepackage{graphicx}
\usepackage{enumitem}
\usepackage{xurl}
\usepackage[export]{adjustbox}
\usepackage{float}
\usepackage{listings}

\usepackage[table,xcdraw,dvipsnames]{xcolor}

\usepackage[backend=biber,style=numeric,sorting=none]{biblatex}

\begin{document}
\title{Case studies of development of verified programs with Dafny for accessibility assessment\thanks{This is an extended version, including the source code, of our FSEN 2023 paper.}}

\titlerunning{Case studies of development of verified programs with Dafny}
%

\author{João Pascoal Faria\inst{1,2}\orcidID{0000-0003-3825-3954} \and
Rui Abreu\inst{1,3}\orcidID{0000-0003-3734-3157}}
\authorrunning{João Pascoal Faria and Rui Abreu}
%
\institute{Faculty of Engineering of the University of Porto, Porto, Portugal \\ 
\email{\{jpf,rma\}@fe.up.pt}
\and
INESC TEC - Institute for Systems and Computer Engineering, Technology and Science, Porto, Portugal \and
INESC ID, Lisbon, Portugal}

\maketitle              

\begin{abstract}
Formal verification techniques aim at formally proving the correctness of a computer program with respect to a formal specification, but the expertise and effort required for applying formal specification and verification techniques and scalability issues have limited their practical application. 
In recent years, the tremendous progress with SAT and SMT solvers enabled the construction of a new generation of tools that promise to make formal verification more accessible for software engineers, by automating most if not all of the verification process. The Dafny system is a prominent example of that trend. However, little evidence exists yet about its accessibility.
To help fill this gap, we conducted a set of 10 case studies of developing verified implementations in Dafny of some real-world algorithms and data structures, to determine its accessibility for software engineers.  
We found that, on average, the amount of code written for specification and verification purposes is of the same order of magnitude as the traditional code written for implementation and testing purposes (ratio of 1.14) -- an ``overhead'' that certainly pays off for high-integrity software. The performance of the Dafny verifier was impressive, with 2.4 proof obligations generated per line of code written, and 24 ms spent per proof obligation generated and verified, on average.
However, we also found that the manual work needed in writing auxiliary verification code may be significant and difficult to predict and master. Hence, further automation and systematization of verification tasks are possible directions for future advances in the field.

\end{abstract}

\section{Introduction}

\subsection{Motivation}

Given the increasing dependence of our society on software-based systems, it is ever more important to assure their correct, secure and safe functioning, particularly for high-integrity systems \cite{boehm2006some}. Since software development is a knowledge-intensive activity and software-based systems are increasingly complex, errors are inevitable, so several techniques need to be applied along the process to catch and fix defects as early as possible.

Testing and reviews are the most widely applied techniques in the software industry for defect detection. However, since ``program testing can be used to show the presence of bugs, but never to show their absence'' \cite{dijkstra1970notes}, testing alone cannot be considered sufficient for high-integrity systems. If properly applied \cite{humphrey2000introduction}, reviews are a cost-effective technique for defect detection and knowledge sharing, but, like with testing, they cannot be used to show the absence of bugs.

By contrast, formal verification techniques aim at formally proving the correctness of a computer program, i.e., show the absence of defects. To that end, we need a formal specification of the program intent and a logic reasoning framework, usually based on Hoare logic \cite{hoare1969axiomatic}. But the expertise and effort required for applying formal specification and verification techniques and scalability issues have limited their practical application.
In recent years, the tremendous progress with SAT and SMT solvers \cite{vardi2016automated}, such as Z3 \cite{moura2008z3}, enabled the construction of a new generation of tools that promise to make formal verification accessible for software engineers, like Dafny  \cite{leino2017accessible}, Frama-C \cite{cuoq2012frama} and Why3 \cite{filliatre2013why3}, by automating most if not all of the verification process. However, little evidence exists yet about their accessibility, regarding the expertise and effort required to apply them.

The authors have used formal specification languages and automated reasoning tools for several years in software engineering research, education, and practice~\cite{abreu2015using,
rebello2012specification,campos2013encoding,diedrich2016applying,lima2020local}.
E.g., in \cite{rebello2012specification}, Alloy \cite{jackson2012software} was used to automatically generate unit tests and mock objects in JUnit\footnote{\url{https://junit.org/}} from algebraic specifications of generic types. 
Although model-based testing approaches such as this one do not guarantee the absence of bugs, they provide a higher assurance than manual test generation and seem to be currently more accessible than formal verification. 

From an educational perspective, the authors are also interested in assessing the feasibility of embedding computer-supported formal specification and verification techniques in undergraduate programs, namely in courses dedicated to studying algorithms and data structures.

\subsection{Objectives and Methodology}

To help fill the gap in the current state of the art regarding accessibility studies, we conducted a set of case studies of developing verified implementations in Dafny of some well-known algorithms and data structures of varying complexity, with the goal of determining its accessibility for software engineering practitioners, students and researchers, with limited training in formal methods.

Table \ref{tab:case_studies} shows the list of case studies. 
They explore formal specification and verification features of increasing complexity. In Sec. \ref{sec:highlights}, we provide some highlights for selected features. For each case study, we collected a few metrics and lessons learned, to help answer our main question, regarding Dafny accessibility. Those metrics and lessons learned are aggregated and discussed in Sec. \ref{sec:results}\. The source code is available in a GitHub repository\footnote{\url{https://github.com/joaopascoalfariafeup/DafnyProjects}} and Appendix \ref{sec:code}.


\begin{table}[ht]
\caption{List of case studies.}
\label{tab:case_studies}
\begin{tabular}{|m{0.21\linewidth}|m{0.77\linewidth}|}
\hline

\textbf{Category} & \textbf{Case study} \\ \hline

\textbf{Numerical algorithms} & 
\begin{itemize}[label=$\circ$,topsep=2pt] 
\item{Integer division (Euclidean division)} 
\item{Natural power of a number (divide and conquer algorithm)} 
\vspace{-8pt} 
\end{itemize} \\  \hline

\textbf{Searching \& sorting algorithms} & 
\begin{itemize}[label=$\circ$,topsep=2pt] 
\item{Binary search} 
\item{Insertion sort} 
\vspace{-8pt} 
\end{itemize}  \\ \hline

\textbf{Collections} &
\begin{itemize}[label=$\circ$,topsep=2pt] 
\item{Priority queue implemented with a binary heap}
\item{Unordered set implemented with a hash table (Hash Set)} 
\item{Ordered set implemented with a binary search tree (Tree Set)} 
\vspace{-8pt} 
\end{itemize} \\ \hline

\textbf{Matching problems} &
\begin{itemize}[label=$\circ$,topsep=2pt] 
\item{Stable marriage problem solved by the Gale-Shapley algorithm} 
\item{Teachers placement problem reduced to stable marriage} 
\vspace{-8pt} 
\end{itemize} \\ \hline

\textbf{Graph algorithms} & 
\begin{itemize}[label=$\circ$,topsep=2pt] \item{Topological sorting (Khan's algorithm \cite{kahn1962topological})} 
\item{Eulerian circuit (Hierholzer's algorithm)}
\vspace{-8pt} 
\end{itemize} \\ \hline

\end{tabular}
\end{table}

\subsection{Structure of the Paper}

Sec. \ref{sec:highlights} presents some highlights about specification and verification features of increasing complexity in the case studies. Sec. \ref{sec:results} consolidates the metrics collected and lessons learned, and draws conclusions regarding our research goal. Related work is discussed in Sec. \ref{sec:relatedwork}. Conclusions and future work are presented in Sec. \ref{sec:conclusions}.

\section{Case Studies Highlights}
\label{sec:highlights}

\subsection{An Introductory Example (Integer Division)}

\begin{figure}[ht]
\begin{center}
\includegraphics[scale=0.7, trim={2cm 16cm 2cm 2.5cm},clip] {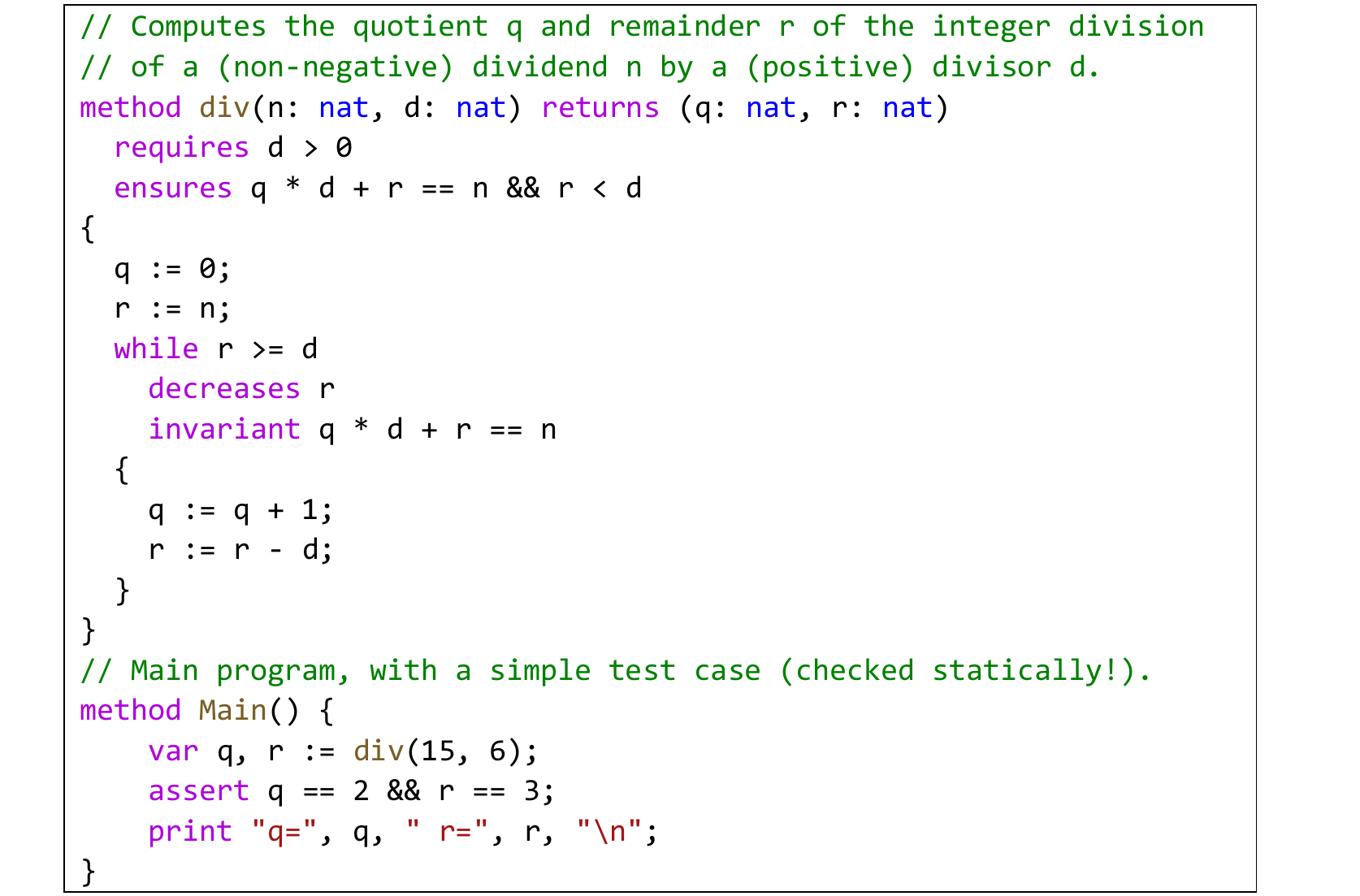}
\end{center}
\caption{A simple program in Dafny for performing integer division.} 
\label{fig:div}
\end{figure}

The self-explanatory program in Fig. \ref{fig:div} explores some basic features of Dafny and serves as our first case study.

Dafny\footnote{\url{https://github.com/dafny-lang/dafny}} \cite{leino2017accessible} is a multi-paradigm programming language and system for the development of verified programs. 
The functional style is typically used for writing specifications, using value types and side-effect-free expressions, functions, and predicates. The procedural and object-oriented styles are typically used for writing implementations, using reference types (arrays, classes, etc.), and methods and statements with side effects.
The Dafny programming system comprises a verifier (based on Z3), compilers that produce code in several target languages (C\#, Java, JavaScript, Go, and C++), and an extension for Visual Studio Code.

The semantics of a method (\texttt{div} in this case) is formally specified by means of pre and postconditions, indicated with the \texttt{requires} and \texttt{ensures} clauses, respectively. 
The Dafny verifier is in charge of checking (with the help of the Z3 theorem prover) if such pre and postconditions are satisfied. When the implementation involves a loop, the user has to provide a loop invariant (with the \texttt{invariant} clause) and, in some cases, a loop variant (with the \texttt{decreases} clause), to help the verifier accomplish its job. 

The \texttt{Main} method is the entry point of a program in Dafny. In this example, it exercises the \texttt{div} method for some inputs, and checks (with \texttt{assert}) and prints the corresponding outputs. Like with pre and postconditions, \texttt{assert} statements are checked statically by the Dafny verifier. In this example, the verifier will try to prove the assertion based only on the postcondition of the \texttt{div} method (i.e., the method body is opaque for this purpose); this makes the verification modular and scalable. Since assertions are checked statically, test cases such as the one shown do actually test the specification in pre-compile time, and not the implementation at run-time; such \textit{static test cases} are useful to detect problems in the specification, e.g., incomplete postconditions.

All the specification constructs and assertions mentioned above (indicated with the \texttt{requires}, \texttt{ensures}, \texttt{invariant}, \texttt{decreases}, and \texttt{assert} clauses) are used as annotations for verification purposes only (during static analysis), but are not compiled into the executable program, so do not cause runtime overhead.

\subsection{Lemmas and Automatic Induction (Power of a Number)} 
\begin{figure}[ht]
\begin{center}
\includegraphics[scale=0.7, trim={2cm 17.5cm 3cm 2.5cm},clip] {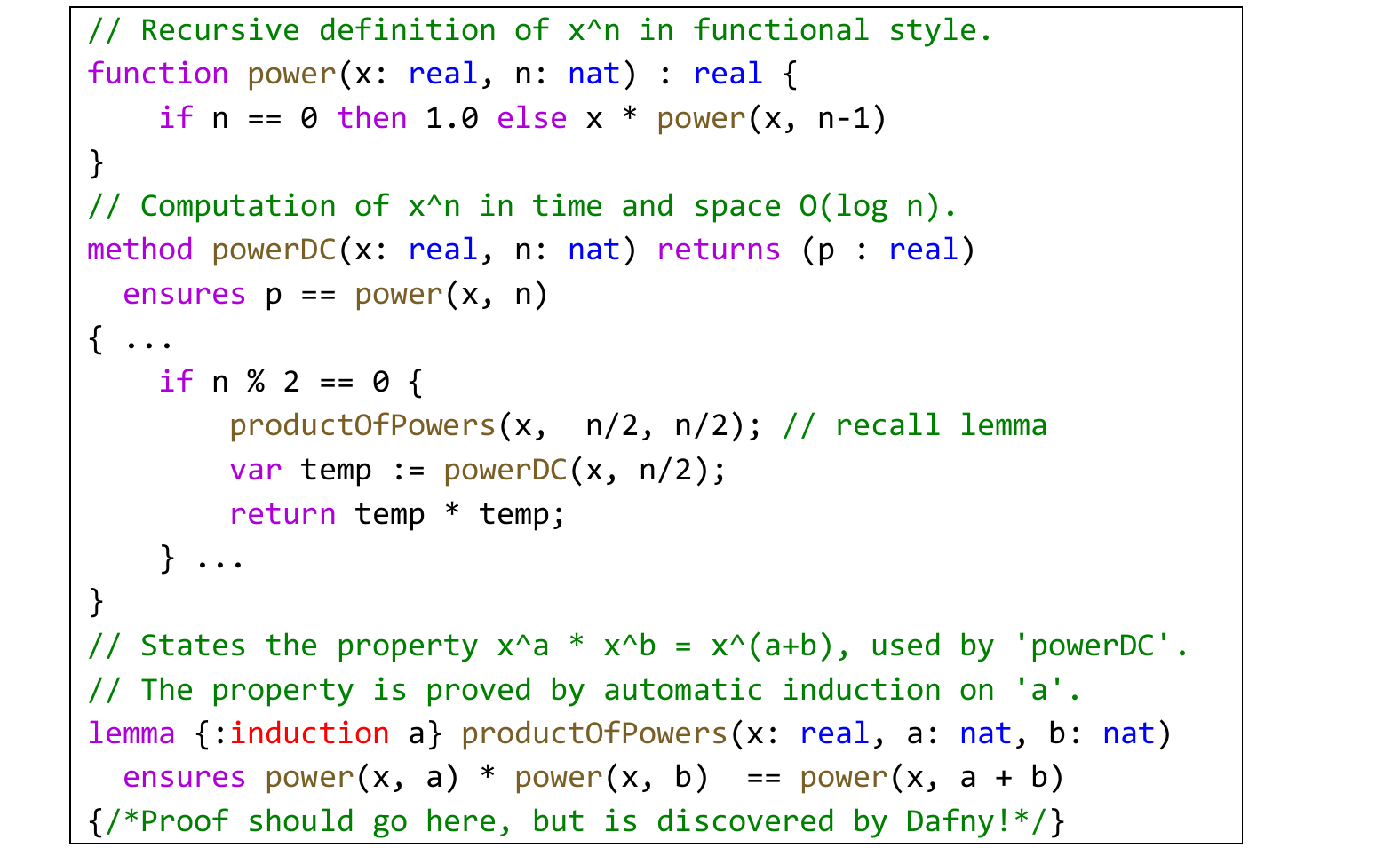}
\end{center}
\caption{Excerpts of a program in Dafny for computing the natural power of a number.} 
\label{fig:power}
\end{figure}

In this case study, the goal is to prove the correctness of a well-known $O(log\ n)$ divide-and-conquer algorithm to compute the natural power of a real number ($x^n$). Self-explanatory excerpts are shown in Fig. \ref{fig:power} and the full code is available in Sec. \ref{sec:power}. It illustrates the usage of lemmas, to specify properties that Dafny alone cannot deduce, and automatic induction, i.e., the ability of Dafny to automatically prove some properties by induction (directive \texttt{:induction a}).

\subsection{Modules, Mutable Objects and Generics (Insertion Sort)} 

\begin{figure}[ht]
\begin{center}
\includegraphics[scale=0.7, trim={2cm 14.8cm 3.5cm 2.5cm},clip] {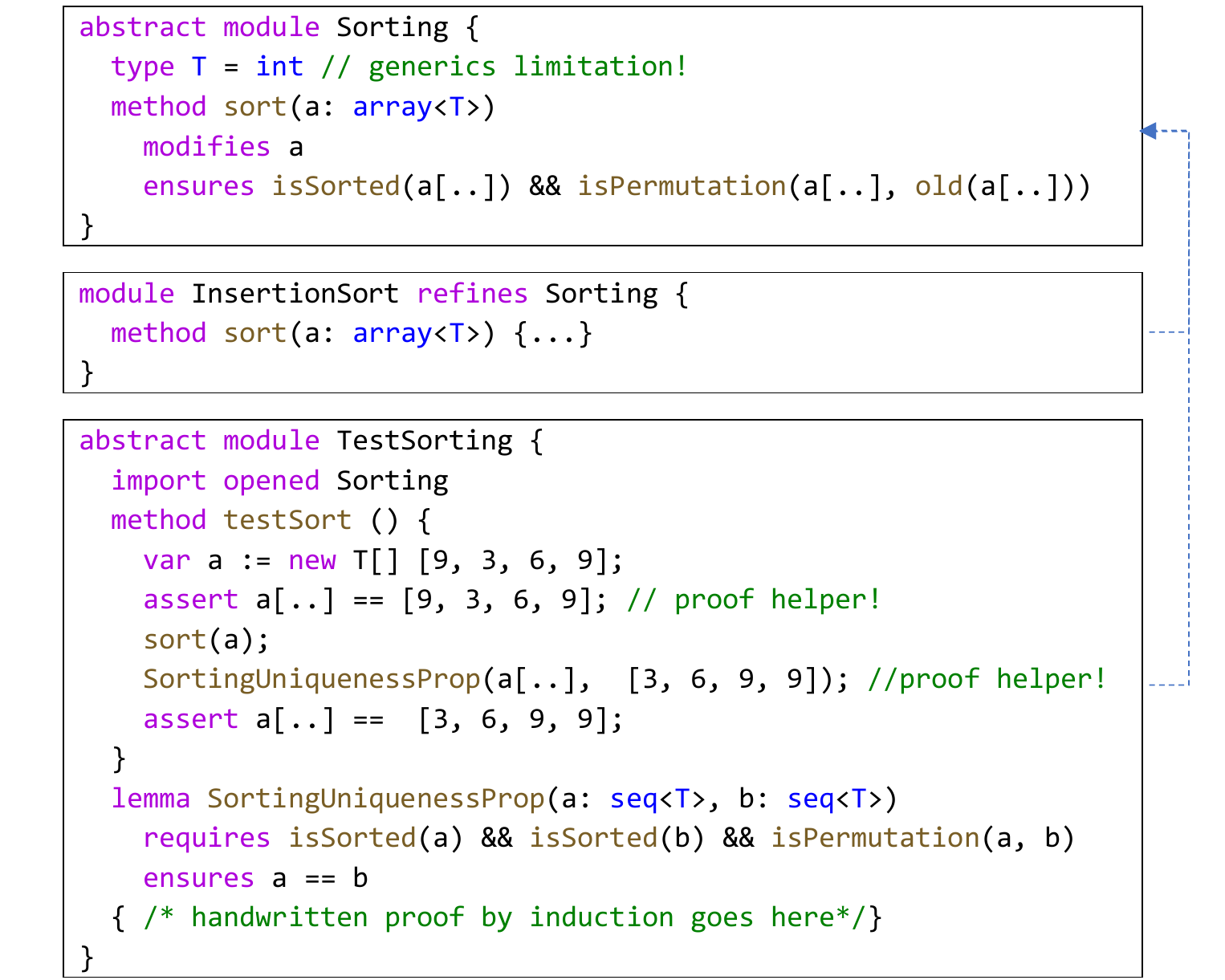}
\end{center}
\caption{Organization of an array sorting program in Dafny using modules.} 
\label{fig:insertionsort}
\end{figure}

In this case study, we explore Dafny features for working with mutable objects (in this case, arrays) and generics, and separating specification, implementation, and test code with modules. Self-explanatory excerpts are shown in Fig. \ref{fig:insertionsort}. 

The array sorting problem is specified by the bodyless \texttt{sort} method in the abstract module \texttt{Sorting}, resorting to auxiliary predicates. The frame condition ``\texttt{modifies a}'' indicates that an implementation may modify the contents referenced by \texttt{a}. In the postcondition, ``\texttt{old(a[...])}'' and ``\texttt{a[..]}'' give the array contents at the begin and end of method execution, respectively, as mathematical sequences. 
Dafny has some support for generic predicates, functions and methods, but, unfortunately, does not support type parameters that are subject to operations other than equality (\texttt{==}); so, for demo purposes, we declared the type of array elements with a specific \texttt{type} definition.   

Sorting algorithms may be provided in concrete modules that refine the abstract module, as in the \texttt{InsertionSort} module, inheriting the method contract and providing the actual algorithm in the body (omitted here). In this case, we just had to provide the loop invariants for the verifier to successfully check the correctness of the insertion sort algorithm with respect to the specification.

The module \texttt{TestSorting} shows an example of a test case of the \texttt{sort} method. For the Dafny verifier to successfully check the test outcome in the last \texttt{assert} statement, we had to write an auxiliary lemma implying that the outcome of \texttt{sort} is unique.
Surprisingly, for the code to be checked successfully, we also had to provide some further ``proof helper'' assertions (as the first assertion) stating trivial facts that we expected to be taken for granted.     

\subsection{State Abstraction and Automatic Contracts (Priority Queue)}

In this case study, we explore Dafny features for separating specification and implementation and handling class invariants in object-oriented programs, following design by contract (DbC) principles. Excerpts of the specification of a priority queue and its implementation with a binary heap are shown in Fig. \ref{fig:insertionsort}. 

The operations’ pre and postconditions of the priority queue (top box in Fig. \ref{fig:insertionsort}) are specified independently of the internal state representation (a binary heap in this case), by resorting to a \textit{state abstraction function} (\texttt{elems}). This function gives the priority queue contents as a multiset (allowing repeated values), and serves only for specification and verification purposes (doesn’t generate executable code); to keep the specification at a high level of abstraction, it doesn’t tell the ordering of elements (which is given by \texttt{deleteMax}).

\begin{figure}[ht]
\begin{center}
    \includegraphics[scale=0.7, trim={2.8cm 9.3cm 2.5cm 2.5cm},clip] {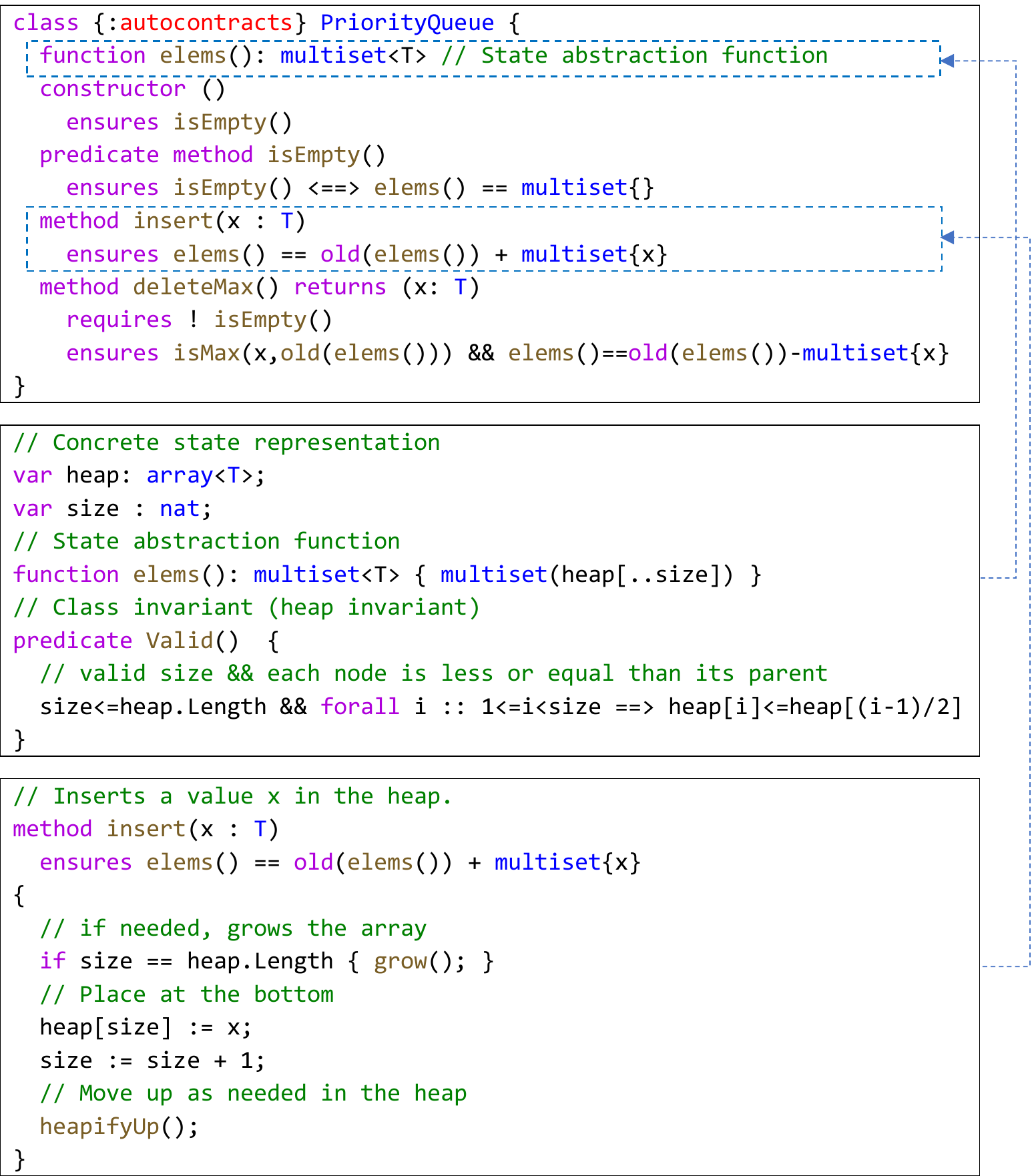}
\end{center}
\caption{Excerpts of a specification (top) of a priority queue and its implementation (center and bottom) with a binary heap in Dafny.} 
\label{fig:priorityqueue}
\end{figure}

In a subsequent refinement (box at the center of Fig. \ref{fig:insertionsort}), it is chosen an internal (concrete) state representation - a binary \texttt{heap} stored in an array. It is also provided an implementation (body) for each method (box at the bottom of Fig. \ref{fig:priorityqueue}). 
The definition and verification of class invariants, stating restrictions on the internal state to be respected at method boundaries, is facilitated in Dafny with so-called automatic contracts, using the ``\texttt{:autocontracts}'' attribute. The class invariant is specified in a predicate \texttt{Valid}; calls to that predicate, together with some frame conditions, are automatically injected in the preconditions of all methods and in the postconditions of all methods and constructors.   

 Thanks to the state abstraction function and the class invariant, the Dafny verifier is able to automatically check the conformity of the methods' implementation (defined in terms of the concrete state) against the methods' pre and postconditons (defined in terms of the abstract state), without further burden from the user! 
 We only had to define an auxiliary lemma, showing that the heap invariant (indicated by the predicate \texttt{Valid} in Fig. \ref{fig:priorityqueue}) implies that the maximum is at the top (array index 0).

\subsection{Proof Techniques (Topological Sorting, Eulerian Circuit)}

\begin{figure}[ht]
\begin{center}
\includegraphics[scale=0.4, trim={1.1cm 3.3cm 4cm 2.2cm},clip] {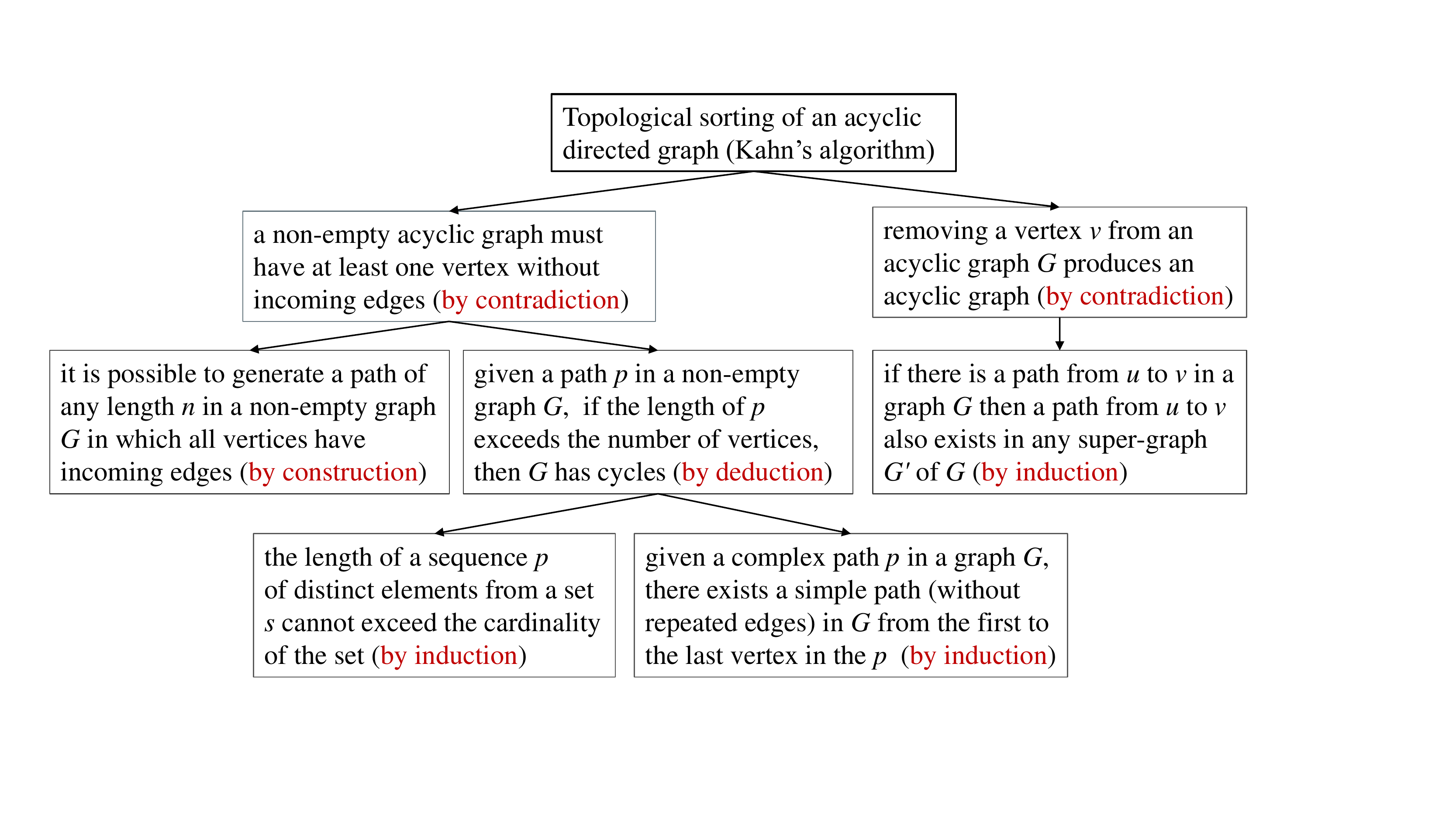}
\end{center}
\caption{Lemmas and proof techniques used to prove the correctness of Kahn's algorithm.} 
\label{fig:proofs}
\end{figure}

Not surprisingly, simple algorithms may require complex proofs, as illustrated in the topological sorting case study. In fact, the Kahn's algorithm \cite{kahn1962topological} can be encoded in just 6 lines of code (at a high level of abstraction), but, to prove its correctness, we had to write 7 auxiliary lemmas, sketched in Fig. \ref{fig:proofs}. Fortunately, Dafny supports a rich variety of proof techniques and is able to fill in most (if not all) of the proof steps, so we only had to provide key intermediate steps, making the handwritten proof of each lemma rather short.   

However, the way the proof steps are written may have a significant impact on the verification time. E.g., in the Eulerian circuit case study, approximately 20 seconds were spent in the verification of a lemma stating that, if an Euler trail $r$ exists in a graph $G$ (i.e., a path that traverses each edge of $G$ exactly once), then each vertex of $G$ has an even number of adjacent vertices, except for the first and last vertex in $r$ in case they are different. The proof is done by induction. By rewriting the inductive step so that the first edge is removed from $r$ and $G$ instead of the last one (possibly better matching the structure of recursive definitions needed in the proof), the verification time was reduced to less than 1 second!

\section{Results and Discussion}
\label{sec:results}

In this section, we summarize the metrics collected and lessons learned from the case studies conducted, and draw some conclusions regarding our research goal.

\subsection{Metrics Collected}

Table \ref{tab:metrics} summarizes the metrics collected in the case studies. 
Size of the code categories described in Table \ref{tab:codecat} is measured in physical lines of code (LOC), ignoring blank lines and comments. 

The execution times were measured in an Intel(R) Core(TM) i7-8750H CPU @ 2.20GHz laptop with 6 cores and 16 GB RAM running Windows 10 Enterprise. We used v2.1.1 of the Dafny extension for VS Code and version 3.3.0 of the Dafny server and, in some cases, version 2.3.0 due to a bug with Z3 and Dafny v3 \footnote{\url{https://github.com/dafny-lang/dafny/issues/1498}}.

\begin{table}[h]
\caption{Results of the case studies (size, time and proof obligations).}
\label{tab:metrics}
\begin{tabular}{|
>{\columncolor[HTML]{FFFFFF}}l |
>{\columncolor[HTML]{FFFFFF}}c |
>{\columncolor[HTML]{FFFFFF}}c |
>{\columncolor[HTML]{FFFFFF}}c |
>{\columncolor[HTML]{FFFFFF}}c |
>{\columncolor[HTML]{FFFFFF}}c |
>{\columncolor[HTML]{FFFFFF}}c |
>{\columncolor[HTML]{FFFFFF}}c |
>{\columncolor[HTML]{FFFFFF}}c |}
\hline
\textbf{Program} &
  \textbf{\begin{tabular}[c]{@{}c@{}}Impl. \\ LOC\end{tabular}} &
  \textbf{\begin{tabular}[c]{@{}c@{}}Test\\ LOC\end{tabular}} &
  \textbf{\begin{tabular}[c]{@{}c@{}}Spec. \\ LOC\end{tabular}} &
  \textbf{\begin{tabular}[c]{@{}c@{}}Verif. \\ LOC\end{tabular}} &
  \textbf{\begin{tabular}[c]{@{}c@{}}Total \\ LOC\end{tabular}} &
  \textbf{\begin{tabular}[c]{@{}c@{}}(S+V)/ \\(I+T)\end{tabular}} &
  \textbf{\begin{tabular}[c]{@{}c@{}}Proof \\ Oblig.\end{tabular}} &
  \textbf{\begin{tabular}[c]{@{}c@{}}Ver.Time\\ (sec.)\end{tabular}} \\ \hline
\textbf{Integer Division}                 & 10           & 5            & 2            & 2            & 19            & 0.27          & 15            & 0.5         \\ 
\textbf{Power of a Number}               & 17           & 7            & 4            & 5            & 33            & 0.38          & 45            & 0.5         \\ 
\textbf{Binary Search}        & 15           & 7            & 7            & 3            & 32            & 0.45          & 51            & 0.5         \\ 
\textbf{Insertion Sort}       & 13           & 13           & 10           & 21           & 57            & 1.19          & 90            & 1           \\ 
\textbf{Priority Queue}       & 74           & 13           & 30           & 35           & 152           & 0.75          & 483           & 3           \\ 
\textbf{Hash Set}             & 86           & 16           & 57           & 38           & 197           & 0.93          & 656           & 16          \\ 
\textbf{Tree Set}             & \textbf{87}  & 13           & 39           & 38           & 177           & 0.77          & \textbf{809}  & 18          \\ 
\textbf{Stable Marriage}     & 50           & \textbf{66}  & 54           & 10           & 180           & 0.55          & 209           & 7           \\ 
\textbf{Topological Sorting} & 19           & 18           & 21           & 94           & 152           & 3.11          & 157           & 3           \\ 
\textbf{Eulerian Circuit}    & 32           & 10           & \textbf{66}  & \textbf{115} & \textbf{223}  & \textbf{4.31} & 407           & \textbf{19} \\ \hline
\textbf{Total}               & \textbf{403} & \textbf{168} & \textbf{290} & \textbf{361} & \textbf{1222} & \textbf{1.14} & \textbf{2922} & \textbf{69} \\ \hline
\end{tabular}
\end{table}

\begin{table}[h]
\caption{Code categories.}
\label{tab:codecat}
\begin{tabular}{|m{0.18\linewidth}|m{0.82\linewidth}|}
\hline
\textbf{Category} & \textbf{Description}  \\ \hline
\textbf{Implemen-tation} & ``Traditional'', compilable, implementation code (method signatures, method bodies, data definitions, etc.).  \\ \hline
\textbf{Test} & Test code (checked statically or dynamically), including assertions.  \\ \hline
\textbf{Specification} & Specification of contracts, including requires and ensures clauses,  class invariants, frame conditions, and auxiliary definitions used in them.  \\ \hline
\textbf{Verification} & Verification helper code, such as, lemmas and all non-compilable code inside method bodies (loop variants, loop invariants, assertions, invocation of lemmas, manipulation of ghost variables, etc.).  \\ 
\hline
\end{tabular}
\end{table}

\begin{figure}[ht]
\begin{center}
\includegraphics[scale=0.4, trim={4cm 9.4cm 21cm 2.2cm},clip] {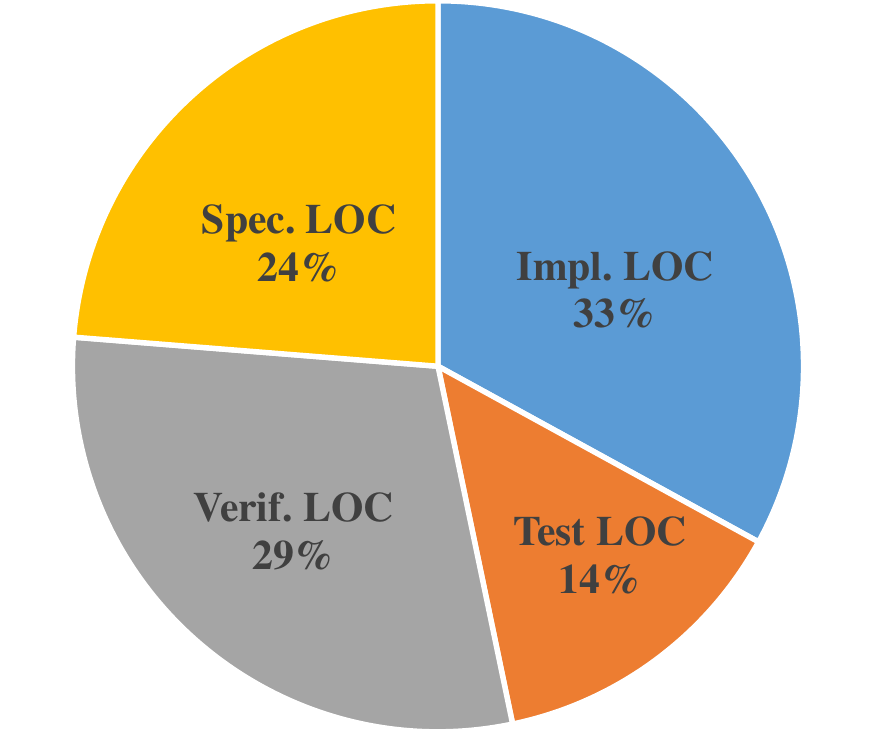}
\end{center}
\caption{Code size (LOC) distribution.} \label{fig:sizedist}
\end{figure}

On average, the amount of code written for formal specification (S) and verification (V) purposes is of the same order of magnitude as the “traditional” code written for implementation (I) and testing (T) purposes -- an ``overhead'' that certainly pays off, at least for high-integrity software. The average ratio is (S+V)/(I+T)=1.14, ranging from 0.27 in the simplest case to 4.31 in the most complex case. 
The pie chart of Fig. \ref{fig:sizedist} shows a balanced size distribution, on average, between the different code categories.

The overhead on user time is difficult to measure as it depends heavily on the user experience. A fair assessment should be done in a different context (in the case studies, the algorithms were known, but the verification strategies had to be discovered in many cases). We believe that, with proper training, in cases where new algorithms have to be designed, the specification and verification effort can be of the same order of magnitude as the design, implementation, and test effort.

The number of proof obligations (POs) generated and checked by the Dafny verifier is impressive, with 2.4 POs generated on average per LOC written (2922 POs/1222 LOC in Table \ref{tab:metrics}), and 7.3 per implementation LOC (2922 POs/403 LOC in Table \ref{tab:metrics}), in the case studies. The performance of the Dafny verifier was also impressive, with 24 ms spent on average per PO generated and verified (69 sec/292 POs in Table \ref{tab:metrics}), in this set of case studies. 


However, based on the experience of the case studies, it is important to note that the verification of some POs may be significantly higher, in the order of minutes, or not even terminate. When that happens, with careful  debugging and refactoring (of assertions, verification code, etc.), one may usually reduce the verification time drastically (as illustrated in the Euler Circuit case study).



\subsection{Lessons Learned}

The lessons learned from the case studies are summarized in Tables \ref{tab:lessons01} and \ref{tab:lessons02}, using a color scheme to highlight \textcolor{PineGreen}{strengths} and \textcolor{BrickRed}{weaknesses}. 
Overall, the Dafny language and verifier proved to be very powerful, automating most of the verification work, with minor language limitations (regarding generics, automatic contracts, and other aspects). Regarding our main research question, the major difficulty we found is that the manual verification work may be significant and difficult to predict and master in non-trivial programs.

\begin{table}[H]
\caption{Lessons learned from the case studies (Part I).}
\label{tab:lessons01}
\begin{tabular}{|m{0.12\linewidth}|m{0.88\linewidth}|}
\hline
\textbf{Category} & \textbf{Lessons learned (\textcolor{PineGreen}{strengths} and \textcolor{BrickRed}{weaknesses})} \\ \hline
\textbf{Dafny Language} & 
\begin{itemize}[topsep=2pt] 
\item \textcolor{PineGreen}{\textbf{Integrated language} for writing \textbf{specifications} (methods’ pre and postconditions), \textbf{implementations} (methods’ bodies), and \textbf{verification helper code} (e.g., loop invariants)[ex: Integer Division].}
\item \textcolor{PineGreen}{Rich set of \textbf{logical quantifiers} (\texttt{forall}, \texttt{exists}, etc.) and \textbf{mathematical collections} (sequences, sets, multisets, maps, etc.), for writing specifications and assertions and describing complex algorithms at a high level of abstraction [ex: Binary Search, Stable Marriage].}
\item \textcolor{PineGreen}{\textbf{Inductive data types} and \textbf{pattern matching} expressions may be used to keep the code at a high level of abstraction [ex: Hash Set].}
\item \textcolor{PineGreen}{\textbf{Null safety}: reference types are not nullable unless they are marked with the ``?'' suffix. [ex: Tree Set]}
\item \textcolor{PineGreen}{Constructs to specify \textbf{frame conditions} and \textbf{query the old object state}, when working with mutable objects  [ex: Insertion Sort].}
\item \textcolor{PineGreen}{\textbf{Modules} enable a clear separation between specification, implementation, and test code [ex: Insertion Sort].
}
\item \textcolor{BrickRed}{Limited support for \textbf{generics}: lack of support for type parameters that are subject to operations other than equality [ex: Binary Search].}
\item \textcolor{BrickRed}{The support for explicitly separating specification and implementation and hiding implementation details in object-oriented programs has room for improvement (e.g., there are no visibility modifiers) [ex: Tree Set].}
\vspace{-8pt} 
\end{itemize} \\ 

 \hline

\textbf{Dafny Compiler} & 
\begin{itemize}[topsep=2pt] 
\item \textcolor{PineGreen}{The Dafny compiler is able to generate executable code in \textbf{multiple target languages} (in this case, only C\# is explored).}
\item \textcolor{PineGreen}{Assertions and other constructs used for specification \& verification purposes are not compiled, so they imply \textbf{no runtime overhead}.}
\vspace{-8pt} 
\end{itemize} \\ 
\hline

\textbf{Dafny Verifier} & 
\begin{itemize}[topsep=2pt] 
\item \textcolor{PineGreen}{In many cases, the verifier is able to \textbf{automatically check that the implementation conforms to the specification}, with minimal user help (that may only have to write loop invariants) [ex: Integer Division].}
\item \textcolor{PineGreen}{Dafny is frequently able to discover loop variants [ex: Binary Search].}
\item \textcolor{PineGreen}{Outside of a method, the method body is opaque for verification purposes (only the pre and postconditions matter), making the verification process modular and scalable.}

\vspace{-8pt} 
\end{itemize} \\ 
\hline

\textbf{Manual Verification Work} & 
\begin{itemize}[topsep=2pt] 
\item \textcolor{PineGreen}{Dafny effectively supports a rich variety of \textbf{proof techniques} (by \textbf{deduction}, by \textbf{induction}, by \textbf{contradiction},  by \textbf{construction},  \textbf{calculational}\cite{leino2013verified}) [ex: Topological Sorting, Tree Set]}
\item \textcolor{BrickRed}{Auxiliary properties may need to be defined by the user (as \textbf{lemmas}) to help the verifier,} \textcolor{PineGreen}{but the proof itself may be greatly or totally automated, with many details automatically filled in;} \textcolor{BrickRed}{discovering what properties need to be defined is not trivial, though [ex: Power, Top. Sort.].}
\item \textcolor{BrickRed}{It is \textbf{difficult to predict when and what manual work will be needed} (beyond writing loop invariants) for a successful verification [ex: Insertion Sort, Topological Sorting].}
\vspace{-8pt} 
\end{itemize} \\ 
\hline

 \end{tabular}
\end{table}

\begin{table}[H]
\caption{Lessons learned from the case studies (Part II).}
\label{tab:lessons02}
\begin{tabular}{|m{0.12\linewidth}|m{0.88\linewidth}|}
\hline
\textbf{Category} & \textbf{Lessons learned 
(\textcolor{PineGreen}{strengths} and \textcolor{BrickRed}{weaknesses})} \\ \hline

\textbf{Auto-matic contracts} & 
\begin{itemize}[topsep=2pt] 
\item \textcolor{PineGreen}{Dafny supports the definition and enforcement of \textbf{class invariants}, especially using the ''\texttt{:autocontracts}`` attribute, also taking care of the generation of appropriate frame conditions [ex: Priority Queue].}
\item \textcolor{BrickRed}{Automatic contracts have room for improvement; in some cases, the user may need to resort to lower level features [ex: Tree Set, Hash Set].}
\item \textcolor{BrickRed}{Getting the contracts right in classes that represent self-referencing data structures may be rather tricky [ex: Tree Set].}
\item \textcolor{BrickRed}{There are apparent conflicts between inheritance and automatic contracts [ex: Priority Queue].}

\vspace{-8pt} 
\end{itemize} \\ 
\hline

\textbf{State Abstraction} & 
\begin{itemize}[topsep=2pt] 
\item \textcolor{PineGreen}{\textbf{State abstraction functions} (ghost functions) allow specifying the semantics (pre/postconditions) of the services provided by a class independently from the implementation (method bodies and internal state representation)  [ex: Priority Queue].}
\item \textcolor{PineGreen}{State abstraction may also be accomplished through \textbf{abstract state variables} (ghost variables), whose abstraction relation to the concrete state variables is specified in the class invariant [ex: Hash Set].}
\vspace{-8pt} 
\end{itemize} \\ 
\hline

\textbf{Testing} & 
\begin{itemize}[topsep=2pt] 
\item \textcolor{PineGreen}{Testing is still relevant, but mainly for \textbf{statically testing the specification}, and not dynamically testing the implementation (proved to be correct with respect to the specification) [ex: Integer division, Ins. Sort].}
\item \textcolor{PineGreen}{Test cases that allow multiple outputs can be easily specified and checked [ex: Insertion Sort].}
\vspace{-8pt} 
\end{itemize} \\ 
\hline

\textbf{Debug-ging and Profiling} & 
\begin{itemize}[topsep=2pt] 
\item \textcolor{PineGreen}{When verification fails, the Dafny language and the Dafny verifier provide several convenient features for debugging purposes, such as the \texttt{assume} statement and the “/tracePOs” option [ex: Eulerian Circuit].}
\item \textcolor{PineGreen}{When the verification time is high, most of the time may be concentrated on one or two assertions. By identifying and rewriting such assertions, the verification time  may be drastically reduced [ex: Eulerian Circuit].}
\vspace{-8pt} 
\end{itemize} \\ 
\hline

 \end{tabular}

\end{table}

\subsection{Accessibility assessment}

We distinguish three levels of competencies required for the development of verified programs in Dafny, with decreasing accessibility:

\begin{itemize}
\item{\textbf{basic}: writing implementation and test code;}
\item{\textbf{intermediate}: writing specifications  (pre/post-conditions, frame conditions, class invariants, and related predicates and functions), and loop variants and invariants;}
\item{\textbf{advanced}: identifying and writing the needed verification code, besides loop variants and invariants (auxiliary lemmas, assertions, ghost variables, etc.). }
\end{itemize}

Lessons learned and metrics collected in the case studies suggest that, even in seemingly simple problems, the user may need to be skilled in advanced verification features and techniques. 

Hence, despite the impressive improvements in automated program verification provided by Dafny, we claim that “we are very close to, but not there yet” regarding the goal of making the development of verified programs accessible for software engineering practitioners and students.
Further automation and systematization of verification tasks (including reusable libraries of common properties and “how to” guides), and integration in mainstream languages, are possible directions for further work in the field. 

Our assessment is corroborated by our experience in teaching a course on ``Formal Methods in Software Engineering''\footnote{\url{https://sigarra.up.pt/feup/en/UCURR_GERAL.FICHA_UC_VIEW?pv_ocorrencia_id=459493}} with 151 master students enrolled in the 2020/21 academic year, with a very positive students feedback (average score of 6 out of 7). 
Students with a high grade ($\ge 85\%$) in a midterm exam  were invited to develop a project in Dafny, consisting in the development of a verified implementation of an algorithm or data structure of medium complexity (hash set, tree set, stable marriage, topological sorting, Eulerian circuit, and text compression). Out of 28 students eligible, 14 picked the challenge, but only 9 delivered, and none met the goals fully. We should note that the classes on formal specification and verification (4 hours per week during 6 weeks) only superficially addressed advanced verification techniques, and the students had a relatively short time to do the project (1 month). 
This experience led us to conclude that more advanced training is required to prepare interested students to handle non-trivial specification and verification problems using Dafny or similar systems.

%

\section{Related Work}
\label{sec:relatedwork}

In \cite{farrell2021using}, the authors report their  experience of using Dafny at the VerifyThis 2021 program verification competition, which aims to evaluate the usability of logic-based program verification tools in a controlled experiment, challenging both the verification tools and the users of those tools. They tackled two of the proposed challenges, and, as a result, identify strengths and weaknesses of Dafny in the verification of relatively complex algorithms. Some strengths mentioned are: Dafny’s ability to prove termination and memory safety with little input; built-in value types, such as sets, sequences, multisets, and maps; predicates and lemmas for more concise specifications; automatic induction; ghost variables and functions. They found it difficult to verify properties of possibly null objects, among other difficulties, impeding them from completing all the tasks on time.

In \cite{furia2015autoproof} the authors argue that formal verification tools are often developed by experts for experts; as a result, their usability by programmers with little formal methods experience may be severely limited. They present their experiences with AutoProof (a tool that can verify the functional correctness of object-oriented software in Eiffel) in two contexts representative of non-expert usage. First, they discuss its usability by students in a graduate course on software verification, who were tasked with verifying implementations of various sorting algorithms. Second, they evaluate its usability in verifying code developed for programming assignments of an undergraduate course. They report their experiences and lessons learned, from which they derive some suggestions for improving the usability of verification tools. They report an average 1.3 ratio between the number of tokens in specification and verification annotations and implementation code, in two small programs. In spite of the differences in context and measurement units, that ratio is of the same order of magnitude as ours.    

In \cite{noble2022more} the authors refer that formal methods are often resisted by students due to perceived difficulty, mathematicity, and practical irrelevance. They redeveloped their software correctness course by taking a programming intensive approach, using Dafny to provide instant formative feedback via automated assessment, which resulted in increased student retention and course evaluation. Although very positive overall, their students found Dafny difficult to learn and use, and the informal observations of the authors are that many of those difficulties stem from ``accidental'' complexity introduced by the Dafny tool. They propose some changes to Dafny's design to tackle some issues found related to program testing, verification debugging, and class invariants, among others.

\section{Conclusions and Future Work}
\label{sec:conclusions}
We conducted a set of case studies of developing verified implementations in Dafny of some real-world and well-known algorithms and data structures, with the goal of determining its accessibility for software engineering students, practitioners and researchers.
We concluded that, despite the impressive improvements in automated program verification provided by Dafny, the manual work needed in writing auxiliary verification code may be significant and difficult to predict and master. Further automation and systematization of verification tasks (including reusable libraries of common properties and “how to” guides), and integration in mainstream languages, are possible directions for further work in the field. We also intend to conduct further studies with other verifiers and problems.
\section*{Acknowledgements} 
This work is financed by National Funds through the Portuguese funding agency, FCT --- Funda\c{c}\~ao para a Ci\^encia e a Tecnologia within project EXPL/CCI-COM/1637/2021.
%
%
%

\printbibliography

@article{boehm2006some,
  title={Some future trends and implications for systems and software engineering processes},
  author={Boehm, Barry},
  journal={Systems Engineering},
  volume={9},
  number={1},
  pages={1--19},
  year={2006},
  publisher={Wiley Online Library}
}

@misc{dijkstra1970notes,
  title={Notes on structured programming},
  author={Dijkstra, Edsger Wybe and others},
  year={1970},
  publisher={Technological University, Department of Mathematics}
}

@book{humphrey2000introduction,
  title={Introduction to the team software process (sm)},
  author={Humphrey, Watts S},
  year={2000},
  publisher={Addison-Wesley Professional}
}

@article{hoare1969axiomatic,
  title={An axiomatic basis for computer programming},
  author={Hoare, Charles Antony Richard},
  journal={Communications of the ACM},
  volume={12},
  number={10},
  pages={576--580},
  year={1969},
  publisher={ACM New York, NY, USA}
}

@article{vardi2016automated,
  title={The automated-reasoning revolution: from theory to practice and back},
  author={Vardi, Moshe Y},
  journal={Distinguished Lecture at NSF CISE, Spring},
  year={2016}
}

@inproceedings{moura2008z3,
  title={Z3: An efficient SMT solver},
  author={Moura, Leonardo de and Bj{\o}rner, Nikolaj},
  booktitle={Int. Conf. on Tools and Algorithms for the Construction and Analysis of Systems},
  pages={337--340},
  year={2008},
  organization={Springer}
}

@article{leino2017accessible,
  title={Accessible software verification with Dafny},
  author={Leino, K Rustan M},
  journal={IEEE Software},
  volume={34},
  number={6},
  pages={94--97},
  year={2017},
  publisher={IEEE}
}

@book{jackson2012software,
  title={Software Abstractions: logic, language, and analysis},
  author={Jackson, Daniel},
  year={2012},
  publisher={MIT press}
}

@inproceedings{rebello2012specification,
  title={Specification-driven unit test generation for java generic classes},
  author={Rebello de Andrade, Francisco and Faria, Joao P and Lopes, Ant{\'o}nia and Paiva, Ana CR},
  booktitle={Int. Conf. on Integrated Formal Methods},
  pages={296--311},
  year={2012},
  organization={Springer}
}

@article{lima2020local,
  title={Local observability and controllability analysis and enforcement in distributed testing with time constraints},
  author={Lima, Bruno and Faria, Jo{\~a}o Pascoal and Hierons, Robert},
  journal={IEEE Access},
  volume={8},
  pages={167172--167191},
  year={2020},
  publisher={IEEE}
}

@inproceedings{leino2013verified,
  title={Verified calculations},
  author={Leino, K Rustan M and Polikarpova, Nadia},
  booktitle={Working Conf. on Verified Software: Theories, Tools, and Experiments},
  pages={170--190},
  year={2013},
  organization={Springer}
}

@article{kahn1962topological,
  title={Topological sorting of large networks},
  author={Kahn, Arthur B},
  journal={Communications of the ACM},
  volume={5},
  number={11},
  pages={558--562},
  year={1962},
  publisher={ACM New York, NY, USA}
}

@inproceedings{cuoq2012frama,
  title={Frama-c},
  author={Cuoq, Pascal and Kirchner, Florent and Kosmatov, Nikolai and Prevosto, Virgile and Signoles, Julien and Yakobowski, Boris},
  booktitle={Int. conf. on software engineering and formal methods},
  pages={233--247},
  year={2012},
  organization={Springer}
}

@inproceedings{filliatre2013why3,
  title={Why3—where programs meet provers},
  author={Filli{\^a}tre, Jean-Christophe and Paskevich, Andrei},
  booktitle={European symposium on programming},
  pages={125--128},
  year={2013},
  organization={Springer}
}

@inproceedings{farrell2021using,
  title={Using dafny to solve the VerifyThis 2021 challenges},
  author={Farrell, Marie and Reynolds, Conor and Monahan, Rosemary},
  booktitle={Proc. of the 23rd ACM Int. Workshop on Formal Techniques for Java-like Programs},
  pages={32--38},
  year={2021}
}

@article{furia2015autoproof,
  title={The AutoProof verifier: Usability by non-experts and on standard code},
  author={Furia, Carlo A and Poskitt, Christopher M and Tschannen, Julian},
  journal={arXiv preprint arXiv:1508.03895},
  year={2015}
}

@inproceedings{noble2022more,
  title={More Programming Than Programming: Teaching Formal Methods in a Software Engineering Programme},
  author={Noble, James and Streader, David and Gariano, Isaac Oscar and Samarakoon, Miniruwani},
  booktitle={NASA Formal Methods Symposium},
  pages={431--450},
  year={2022},
  organization={Springer}
}

@article{abreu2015using,
  title={Using constraints to diagnose faulty spreadsheets},
  author={Abreu, Rui and Hofer, Birgit and Perez, Alexandre and Wotawa, Franz},
  journal={Software Quality Journal},
  volume={23},
  number={2},
  pages={297--322},
  year={2015},
  publisher={Springer}
}

@inproceedings{campos2013encoding,
  title={Encoding test requirements as constraints for test suite minimization},
  author={Campos, Jos{\'e} and Abreu, Rui},
  booktitle={2013 10th Int. Conf. on Information Technology: New Generations},
  pages={317--322},
  year={2013},
  organization={IEEE}
}

@inproceedings{diedrich2016applying,
  title={Applying simulated annealing to problems in model-based diagnosis},
  author={Diedrich, Alexander and Feldman, Alexander and Perdomo-Ortiz, Alejandro and Abreu, Rui and Niggemann, Oliver and de Kleer, Johan},
  booktitle={Int. Workshop on Principles of Diagnosis: DX-2016},
  number={ARC-E-DAA-TN35662},
  year={2016},
  organization={ebook DX conference series}
}

\appendix
\section{Code of the Case Studies}
\label{sec:code}

\subsection{Integer Division}
\small
\begin{lstlisting}
/*
* The Dafny "Hello, World!": a simple program for performing
* integer division.
*/

// Computes the quotient 'q' and remainder 'r' of  the integer
// division of a (non-negative) dividend 'n' by a (positive)
// divisor 'd'.
method div(n: nat, d: nat) returns (q: nat, r: nat)
  requires d > 0
  ensures q * d + r == n && r < d
{
  q := 0; 
  r := n;  
  while r >= d
    decreases r
    invariant q * d + r == n
  {
    q := q + 1;
    r := r - d;
  }
}

// Main program, with a simple test case (checked statically!)
method Main() {
    var q, r := div(15, 6);
    assert q == 2 && r == 3;
    print "q = ", q, " r=", r, "\n";
}
\end{lstlisting}

\subsection{Power of a Number}
\label{sec:power}
\small
\begin{lstlisting}
/* 
* Formal verification of an O(log n) algorithm to calculate 
* the natural power of a real number (x^n), illustrating the
* usage of lemmas and automatic induction in Dafny.
*/

// Recursive definition of x^n in functional style, 
// with time and space complexity O(n).
function power(x: real, n: nat) : real {
    if n == 0 then 1.0 else x * power(x, n-1)
}

// Computation of x^n in time and space O(log n).
method powerDC(x: real, n: nat) returns (p : real)
  ensures p == power(x, n)
{
    if n == 0 {
        return 1.0;
    }
    else if n == 1 {
        return x;
    }
    else if n % 2 == 0 {
        productOfPowers(x,  n/2, n/2); // recall lemma
        var temp := powerDC(x, n/2);
        return temp * temp;
    }
    else {
        productOfPowers(x, (n-1)/2, (n-1)/2); // recall lemma  
        var temp := powerDC(x, (n-1)/2);
        return temp * temp * x;
    } 
}

// States the property x^a * x^b = x^(a+b). 
// The property is proved by automatic induction on 'a'.
lemma {:induction a} productOfPowers(x: real, a: nat, b: nat) 
  ensures power(x, a) * power(x, b)  == power(x, a + b) 
{ }


// A few test cases (checked statically by Dafny).
method testPowerDC() {
    var p1 := powerDC( 2.0, 5); assert p1 == 32.0;
    var p2 := powerDC(-2.0, 2); assert p2 == 4.0;
    var p3 := powerDC(-2.0, 1); assert p3 == -2.0;
    var p4 := powerDC(-2.0, 0); assert p4 == 1.0;
    var p5 := powerDC( 0.0, 0); assert p5 == 1.0;
}
\end{lstlisting}

\subsection{Binary Search}
\small
\begin{lstlisting}
/* 
 * Formal verification of the binary search algorithm with 
 * Dafny.
 */

type T = int // for demo purposes, but could be another type 

// Checks if array 'a' is sorted.
predicate isSorted(a: array<T>)
  reads a
{
  forall i, j :: 0 <= i < j < a.Length ==> a[i] <= a[j]
}
  
// Finds a value 'x' in a sorted array 'a', and returns 
// its index, or -1 if not found. 
method binarySearch(a: array<T>, x: T) returns (index: int)
  requires isSorted(a)
  ensures (0 <= index < a.Length && a[index] == x) 
        || (index == -1 && x !in a[..])
{   
  var low, high := 0, a.Length;
  while low < high 
    invariant 0 <= low <= high <= a.Length
    invariant x !in a[..low] && x !in a[high..]
  {
    var mid := low + (high - low) / 2;
    if {
      case a[mid]  < x => low := mid + 1;
      case a[mid]  > x => high := mid;
      case a[mid] == x => return mid;
    }
  }
  return -1;
}

// Simple test cases to check the post-condition.
method testBinarySearch() {
  var a := new int[5] [1, 4, 4, 6, 8];
  assert a[..]  == [1, 4, 4, 6, 8]; // Proof helper
  var id1 := binarySearch(a, 6); assert id1 == 3;
  var id2 := binarySearch(a, 3); assert id2 == -1;
  var id3 := binarySearch(a, 4); assert id3 in {1, 2};
} 
\end{lstlisting}

\subsection{Insertion Sort}
\small
\begin{lstlisting}
/* 
 * Formal verification of the insertion sort algorithm 
 * with Dafny. 
 */

 // Contract for sorting algorithms.
abstract module Sorting {
  type T = int // for demo purposes, but could be another type

  // Abstract method defining the contract (semantics) of
  // array sorting.
  method sort(a: array<T>)
    modifies a    
    ensures isSorted(a[..])
    ensures multiset(a[..]) == multiset(old(a[..]))
  // Auxiliary predicate that checks if a sequence 'a' 
  // is sorted. 
  predicate isSorted(s: seq<T>) {
    forall i, j :: 0 <= i < j < |s| ==> s[i] <= s[j] 
  }
}

 // Static tests of the Sorting contract 
abstract module TestSorting {
  import opened Sorting  

  method testSortSimple() {
    var a := new T[] [9, 4, 6, 3, 8]; 
    assert a[..] == [9, 4, 6, 3, 8]; // prover helper!
    sort(a);
    assert a[..] == [3, 4, 6, 8, 9];
  }  

  method testSortWithDups() {
    var a := new T[] [9, 3, 6, 9];
    assert a[..] == [9, 3, 6, 9]; // prover helper
    sort(a);
    SortingUniquenessProp(a[..],  [3, 6, 9, 9]);
    assert a[..] ==  [3, 6, 9, 9]; // assertion violation (!?)
  }

  // State and prove by induction the property that, if two
  // sequences are sorted and have the same multiset of
  // elements, then they must be identical (so sorting has a
  // unique solution). 
  lemma SortingUniquenessProp(a: seq<T>, b: seq<T>)
    requires isSorted(a) && isSorted(b) 
             && multiset(a) == multiset(b) 
    ensures a == b
  {
    // recalls useful properties about sequences and their 
    // multisets  
    seqProps(a);
    seqProps(b);

    // key steps of proof by induction on 'a' and 'b' 
    // (the rest is filled in by Dafny) 
    if |a| > 0 {
      SortingUniquenessProp(a[1..], b[1..]);
    }
  }

  // States two properties about sequences (proved by Dafny
  // alone): 
  // - sequence concatenation reverts splitting in head and
  //   tail;
  // - elements of a sequence belong to its multiset.
  lemma seqProps(a: seq<T>) 
    ensures |a| > 0 ==> a == [a[0]] + a[1..] 
    ensures forall i :: 0 <= i < |a| ==> a[i] in multiset(a)
  {}
}

module InsertionSort refines Sorting {
  // Sorts array 'a' using the insertion sort algorithm.
  // Inherits the contract from Sorting.
  method sort(a: array<T>) {
    for i := 0 to a.Length
      invariant isSorted(a[..i]) 
      invariant multiset(a[..]) == multiset(old(a[..]))
    {
      var j := i;
      while j > 0 && a[j-1] > a[j]
        invariant forall l, r :: 0 <= l < r <= i && r != j 
                     ==> a[l] <= a[r] 
        invariant multiset(a[..]) == multiset(old(a[..]))
      {
        a[j-1], a[j] := a[j], a[j-1]; //swap(parallel assign.)
        j := j - 1;
      }
    }
  }
}
\end{lstlisting}

\subsection{Priority Queue}
\small
\begin{lstlisting}
/* 
* Formal specification and verification of a Priority Queue
* implemented as a heap. A heap is a partially ordered set
* represented in an array, suited to implement priority 
* queues operations insert and deleteMax in O(heapSize).
* Illustrates the verification of object-oriented programs
*/
 
type T = int // for demo purposes, but could be real, etc.



class {:autocontracts} PriorityQueue {
 
  // Concrete state representation
  var heap: array<T>;
  var size : nat;
 
  // Configuration parameters
  static const initialCapacity := 10; 
 
  // Class invariant (heap invariant + automatic things
  // generated by :autocontracts)
  predicate Valid()  {
    heapInv()
  }
 
  // Heap invariant
  predicate {:autocontracts false} heapInv()
     reads this, heap
  {
    // valid size 
    size <= heap.Length
    // each node is less or equal than its parent
    && forall i :: 1 <= i < size ==> heap[i] <= heap[(i-1)/2]
  }
 
  // State abstraction function: gets the heap contents as a
  // multiset.
  function elems(): multiset<T> 
  { multiset(heap[..size]) }
 
  // Initializes the heap as empty.
  constructor()
    ensures isEmpty()
  {
    heap := new T[initialCapacity];
    size := 0; 
  }
 
  // Checks if the heap is empty
  predicate method isEmpty() 
    ensures isEmpty() <==> elems() == multiset{}
  {
    // to help proving the post-condition
    assert elems() == multiset{} <==> |elems()| == 0;
    // actual expression 
    size == 0
  }  
 
  // Inserts a value x in the heap.
  method insert(x : T)
    ensures elems() == old(elems()) + multiset{x}
  {
    if size == heap.Length {
      grow();
    }
    // Place at the bottom
    heap[size] := x;
    size := size + 1;
    // Move up as needed in the heap
    heapifyUp();
  }
 
  // Method used internally to grow the heap capacity 
  method grow()
    requires size == heap.Length 
    ensures heap.Length > size
    ensures heap[..size] == old(heap[..size])
  {
      var oldHeap := heap;
      heap := new T[if size == 0 then initialCapacity 
                    else 2 * size];
      forall i | 0 <= i < oldHeap.Length {
        heap[i] := oldHeap[i];
      }
  }
 
 
  // Auxiliary method to move a dirty node from the bottom
  // upwards in the heap
  method {:autocontracts false} heapifyUp()
   requires size > 0 && heapifyUpInv(size-1) 
   modifies heap
   ensures heapInv() 
   ensures multiset(heap[..size])==old(multiset(heap[..size]))
  {
    var k := size - 1;
    while k > 0 && heap[k] > heap[(k - 1) / 2]
      invariant 0 <= k < size
      invariant heapifyUpInv(k) 
      invariant multiset(heap[..size]) ==
                old(multiset(heap[..size]))
    {
      heap[k], heap[(k-1) / 2] := heap[(k - 1) / 2], heap[k];
      k := (k - 1) / 2;
    }
  }
 
  // During heapifyUp, while moving a node up at index k,
  // there are some differences:
  // children of k are sorted wrt parent of k, and k is not
  // sorted wrt its parent.
  predicate {:autocontracts false} heapifyUpInv(k: nat)
    reads this, heap
  {
    size <= heap.Length 
    && (forall i :: 1 <= i < size 
                    && i != k ==> heap[i] <= heap[(i - 1)/2])
    && (k > 0 ==> forall i :: 1 <= i < size && (i-1)/2 == k
                   ==> heap[i] <= heap[((i - 1)/2 - 1)/2])
  }
 
  // Deletes and retrieves the maximum value in the heap 
  // (assumed not empty).
  method deleteMax() returns (x: T)
    requires ! isEmpty()
    ensures isMax(x, old(elems()))
    ensures elems() == old(elems()) - multiset{x}
  {
    // recall the lemma ...  
    maxIsAtTop(); 
    // pick the maximum from the top
    x := heap[0];  
    // reduce the size
    size := size - 1; 
    if size > 0 {
      // move last element to top
      heap[0] := heap[size]; 
      // move down as needed in the heap
      heapifyDown(); 
    }
  }

  // Deletes and retrieves the maximum value in the heap
  // (assumed not empty).
  method geteMax() returns (x: T)
    requires ! isEmpty()
    ensures isMax(x,elems())
  {
    maxIsAtTop(); 
    return heap[0];  
  }
 

  // Auxiliary predicate to check if a value is a maximum in
  // a multiset.
  predicate isMax(x: T, m: multiset<T>) {
    x in m && forall y :: y in m ==> y <= x
  }
 
  // Auxiliary method to move a dirty node from the top down
  // in the heap
  method {:autocontracts false} heapifyDown() 
    requires size > 0 && heapifyDownInv(0) 
    modifies heap
    ensures heapInv() 
    ensures multiset(heap[..size]) ==
            old(multiset(heap[..size]))
  {
    var k := 0;
    while true 
      decreases size - k
      invariant 0 <= k < size
      invariant heapifyDownInv(k) 
      invariant multiset(heap[..size]) == 
                old(multiset(heap[..size]))
    {
      var leftChild := 2 * k + 1; // index of left child
      var rightChild := 2 * k + 2;
      if leftChild >= size {               
        return;  // reached the bottom
      }
      var maxChild := if rightChild < size 
                      && heap[rightChild] > heap[leftChild] 
                      then rightChild else leftChild;
      if heap[k] > heap[maxChild] {                
        return; // already sorted
      }
      // move up and continue
      heap[k], heap[maxChild] := heap[maxChild], heap[k];
      k := maxChild;
    }
  }
 
  // During heapifyDown, while moving a node down at index k,
  // there are some differences:
  // children of k are sorted wrt parent of k, and k is not
  // sorted wrt its children.
  predicate {:autocontracts false} heapifyDownInv(k: nat)
    reads this, heap
  {
    size <= heap.Length
    && (forall i :: 1 <= i < size && (i-1)/2 != k 
          ==> heap[i] <= heap[(i - 1)/2])
    && (k > 0 ==> forall i :: 1 <= i < size && (i-1)/2 == k 
                   ==> heap[i] <= heap[((i - 1)/2 - 1)/2])
  }
  
  // Lemma stating that the maximum is at the top of the heap
  // (position 0). This property is assumed by deleteMax and
  // follows from the heap invariant.
  // Proved by induction on the size of the heap, reason why
  // it receives a parameter with the size to consider.
  lemma {:induction n} maxIsAtTop(n: nat := size)
    requires n <= size
    ensures forall i :: 0 <= i < n ==> heap[i] <= heap[0]
  {}
}
 
// A simple test scenario.
method testPriorityQueue() {
  var h := new PriorityQueue();
  assert h.isEmpty();
  h.insert(2);
  h.insert(5);
  h.insert(1);
  h.insert(1);
  var x := h.deleteMax(); assert x == 5;
  x := h.deleteMax(); assert x == 2;
  x := h.deleteMax(); assert x == 1;
  x := h.deleteMax(); assert x == 1;
  assert h.isEmpty();  
}
\end{lstlisting}

\subsection{Hash Set}
\small
\begin{lstlisting}
/*
 * Verified implementation of a hash set with open addressing
 * and linear probing in Dafny.
 * Provides the fundamental set operations (contains, insert,
 * delete), specified at an abstract level, resorting to an
 * abstract state variables 'elems' with the set contents. 
 */

// Datatype for the content of each cell of the hash table.
// It stores a value of type T, Nil (no value) or Deleted
// (cell marked as deleted).
datatype Cell<T> = Nil | Deleted | Some(value: T) 

// Function type for hash functions 
type HashFunction<!T> = (T) -> nat 

// Represents a hash set of elements of type T (comparable for
// equality), i.e., a set stored in a hash table. 
// Uses the "autocontracts" attribute to automatically inject
// class invariant checking and frame conditions
// in methods' pre and post-conditions.
class {:autocontracts} HashSet<T(==)> {

  // Ghost variable (abstract state variable) used for
  // specification purposes only.
  ghost var elems : set<T>;

  // Concrete state variable with internal representation. 
  var hashTable: array<Cell<T>>; 

  // Hash function to be used (provided to the constructor). 
  const hash: HashFunction<T>;

  // Number of positions used (with some value) and marked as
  // deleted in the hash table. 
  var used: nat;
  var deleted: nat;

  // Initial capacity of the hash table.
  static const initialCapacity := 101;

  // Ghost predicate that formalizes the class invariant.
  predicate Valid()  {
    // Constraint that define the abstraction relation between
    // abstract and concrete state variables
    elems == valSet(hashTable, hashTable.Length) 
    // Constraints on the internal state representation
    && hashTable.Length > 0 
    && hashTableInv(hashTable)
    && used == |valSet(hashTable, hashTable.Length)|
    && deleted == |delSet(hashTable, hashTable.Length)|
  }

  // Ghost predicate that checks the consistency of a hash
  // table 't'. 
  predicate {:autocontracts false} 
  hashTableInv(t: array<Cell<T>>)
    reads t 
  { 
    forall i :: 0 <= i < t.Length && t[i].Some? 
           ==> validPos(t[i].value, i, t)
  }

  // Ghost predicate that checks that 'i' is a valid position
  // for value 'x' in hash table 't'.
  // ('x' may be or not currently stored in that position) 
  predicate {:autocontracts false} 
  validPos(x: T, i: nat, t: array<Cell<T>>)
    requires 0 <= i < t.Length
    reads t 
  {
    var h := hash(x) % t.Length;
    h == i  
    || (h < i && forall j :: h <= j < i 
                    ==> t[j] != Nil && t[j] != Some(x))
    || (h > i && forall j :: h <= j < t.Length || 0 <= j < i
                    ==> t[j] != Nil && t[j] != Some(x))
  }

  // Ghost function that retrieves the set of values stored in
  // the first 'n' positions of hash table 't'. 
  function {:autocontracts false} 
  valSet(t: array<Cell<T>>, n: nat): set<T>
    requires 0 <= n <= t.Length
    reads t
  { set i | 0 <= i < n && t[i].Some? :: t[i].value }

  // Ghost function that retrieves the set of positions marked
  // as Deleted in the first 'n' positions of hash table 't'. 
  function {:autocontracts false} 
  delSet(t: array<Cell<T>>, n: nat): set<nat> 
    requires 0 <= n <= t.Length
    reads t
  { set i | 0 <= i < n && t[i].Deleted? }

  // Ghost function that retrieves the set of positions marked
  // as Nil in the first 'n' positions of hash table 't'. 
  function {:autocontracts false} 
  nilSet(t: array<Cell<T>>, n: nat): set<nat> 
    requires 0 <= n <= t.Length
    reads t
  { set i | 0 <= i < n && t[i].Nil? }

  // Auxiliary lemma that states the following property: the
  // sum of the sizes of valSet, delSet and nilSet
  // of a valid hash table is equal to the length of the hash
  // table (array). 
  // This is true because the hash table invariant implies
  // that there are no duplicate values stored. 
  // The proof is done by induction on the length of the table
  // (omitting steps filled in by Dafny).
  lemma {:autocontracts false} 
  countingLemma(ht: array<Cell<T>>, len: nat, v: nat, d: nat,
                n: nat)
    requires 0 <= len <= ht.Length
    requires hashTableInv(ht)
    requires v == |valSet(ht, len)| 
             && d == |delSet(ht, len)| 
             && n == |nilSet(ht, len)|
    ensures v + d + n == len
  {
     if len > 0 {
        var vs := valSet(ht, len);
        var ds := delSet(ht, len);
        var ns := nilSet(ht, len); 
        var vs1 := valSet(ht, len-1);
        var ds1 := delSet(ht, len-1);
        var ns1 := nilSet(ht, len-1); 
        // recursive part
        countingLemma(ht, len-1, |vs1|, |ds1|, |ns1|);
        // incremental part
        match ht[len-1] {
          case Deleted => assert vs == vs1 
                          && ds == ds1 + {len-1} && ns == ns1;
          case Nil     => assert vs == vs1 
                          && ds == ds1 && ns == ns1 + {len-1};
          case Some(x) => assert vs == vs1 + {x} 
                          && ds == ds1 && ns == ns1;
        }
     }
  }

  // Internal predicate that checks if the hash table is
  // 'full', in the sense that all positions are occupied with
  // a value or are marked as deleted (i.e., there are no
  // positions with Nil). In that case, inserting a new value
  // might not be possible. 
  predicate method full()
   ensures full() <==> nilSet(hashTable,hashTable.Length)=={} 
  {
    // to help proving the post-condition (equivalence):  
    countingLemma(hashTable, hashTable.Length, used, deleted,
                  |nilSet(hashTable, hashTable.Length)|);
    // the actual function value
    used + deleted == hashTable.Length
  } 
  
  // Public method that checks if this set contains a value x.
  method contains(x: T) returns (res: bool)
    ensures res <==> x in elems
  {
    var pos := locate(x);
    return pos != -1 && hashTable[pos] == Some(x);
  }

  // Internal method that determines the location ('pos') for
  // a value 'x' (existent or to be inserted).
  // If such a location cannot be found (because the table is
  // full), returns -1.
  // In the case of a new value, tries to reuse positions
  // marked as deleted.
  method locate(x: T) returns (pos: int)
    requires Valid()
    ensures x in elems ==> 0 <= pos < hashTable.Length 
                           && hashTable[pos] == Some(x)
    ensures x !in elems ==> (pos == -1 && full()) || 
                            (0 <= pos < hashTable.Length 
                            && !hashTable[pos].Some? 
                            && validPos(x, pos, hashTable))
  {
     var h := hash(x) % hashTable.Length;
     var reuse := -1;
     for i := h to hashTable.Length 
       invariant forall j :: h <= j < i 
           ==> hashTable[j] != Nil && hashTable[j] != Some(x)
       invariant reuse == -1 || 
            (h <= reuse < i && hashTable[reuse] == Deleted)  
     {
        if hashTable[i] == Nil || hashTable[i] == Some(x) {
          return i;
        }
        if hashTable[i] == Deleted && reuse == -1 {
           reuse := i;
        }
     }
     for i := 0 to h 
       invariant forall j :: 0<=j<i || h<=j<hashTable.Length 
            ==> hashTable[j] != Nil && hashTable[j] != Some(x)
       invariant reuse == -1 
        || ((0 <= reuse < i || h <= reuse < hashTable.Length) 
            && hashTable[reuse] == Deleted)
     {
        if hashTable[i] == Nil || hashTable[i] == Some(x) {
          return i;
        }
        if hashTable[i] == Deleted && reuse == -1 {
           reuse := i;
        }
     }
     return reuse;
  }

  // Public constructor that receives the hash function to be
  // used and initializes the set as empty.
  constructor (hash: HashFunction<T>)
    ensures elems == {}
  {
    // initialize concrete state variables 
    this.hash := hash;
    hashTable := new Cell<T>[initialCapacity] (_ => Nil);
    used := 0;
    deleted := 0;
    // initialize ghost/abstract state variables 
    elems := {};
  }

  // Internal method that inserts a new value 'x' into the
  // hash set, guaranteed to be not full.
  method insertAux(x : T)
    requires x !in elems
    requires ! full()
    ensures elems == old(elems) + {x}
    ensures deleted <= old(deleted) // usefull for rehash?
    ensures hashTable == old(hashTable) // usefull for rehash?
 {
    var i := locate(x);
    if hashTable[i] == Deleted {
      // to help proving that deleted > 0
      assert i in delSet(hashTable, hashTable.Length);

      // now, can decrement
      deleted := deleted - 1;
    } 
    hashTable[i] := Some(x);
    used := used + 1;
    elems := elems + {x};

    // to help proving that elems == valSet()  
    assert forall k :: 0 <= k < hashTable.Length && k != i 
               ==> hashTable[k] == old(hashTable[k]);
    
    // to help proving that deleted == |delSet()|
    assert delSet(hashTable, hashTable.Length) ==
             old(delSet(hashTable, hashTable.Length)) - {i};
  }

  // Internal method that grows and cleans up the hash table.
  method rehash()
   ensures ! full()
   ensures elems == old(elems)
  {
    var oldTable := hashTable;

    hashTable:= new Cell<T>[hashTable.Length*2+1] (_ => Nil); 
    deleted := 0;
    used := 0;
    elems := {};    
    // need also to update ghost variable 'Repr' generated by
    // :autocontracts, to be able to call InsertAux in a valid
    // state
    Repr := {this, hashTable};

    for i := 0 to oldTable.Length
      invariant elems == valSet(oldTable, i)
      invariant deleted == 0 // to prove !full()
      invariant oldTable !in Repr 
          // to assure is not changed by insertAux
      invariant Valid() 
          // to ensure consistency is maintained
      invariant oldTable.Length < hashTable.Length 
          // to prove !full()
      invariant fresh(Repr - old(Repr)) 
          // to enable insertAux to modify the new hashTable
      invariant oldTable.Length < hashTable.Length 
          // to prove !full()
      invariant forall k :: 0 <= k < oldTable.Length 
                 ==> oldTable[k] == old(hashTable[k]) 
          // to prove post-condition
    {
      // to help proving ! full()
      countingLemma(oldTable, i, |valSet(oldTable, i)|, 
                |delSet(oldTable, i)|, |nilSet(oldTable, i)|);

      if (oldTable[i].Some?) {
        insertAux(oldTable[i].value);
      }
    }

    // to help proving ! full()
    var i := oldTable.Length;
    countingLemma(oldTable, i, |valSet(oldTable, i)|,
                |delSet(oldTable, i)|, |nilSet(oldTable, i)|);
  }

  // Inserts a new value 'x' into this hash set.
  method insert(x : T)
    requires x !in elems
    ensures elems == old(elems) + {x}
  {
    if full() {
       rehash();
    }
    insertAux(x);
  }

  // Deletes an existent value 'x' from this hash set.
  method delete(x : T)
    requires x in elems
    ensures elems == old(elems) - {x}
 {
    var h := hash(x) % hashTable.Length;
    var i := locate(x);
    elems := elems - {x};
    hashTable[i] := Deleted;
    deleted := deleted + 1;
    used := used - 1;

    // to help proving that elems == valSet(hashTable,
    // hashTable.Length)  
    assert forall k :: 0 <= k < hashTable.Length && k != i 
            ==> hashTable[k] == old(hashTable[k]);

    // to help proving that deleted == |delSet()|
    assert delSet(hashTable, hashTable.Length) ==
            old(delSet(hashTable, hashTable.Length)) + {i};
  }
}

method testHashSet() {
  var h := new HashSet<string>(x => |x|);
  assert h.elems == {};
  h.insert("Hello");
  assert h.elems == {"Hello"};
  h.insert("World");
  assert h.elems == {"Hello", "World"};
  var found := h.contains("Hello");
  assert found;
  found := h.contains("ANSI");
  assert !found;
  h.delete("Hello");
  assert h.elems == {"World"};
  found := h.contains("Hello");
  assert !found;
}
\end{lstlisting}

\subsection{Tree Set}
\small
\begin{lstlisting}
/* 
* Specification and verification of a sorted set implemented
* with a binary search tree (BST).
* Illustrates the usage of ghost variables for data
* abstraction and separation of specification and
* implementation.
* Uses the {:autocontracts} attribute to take care of class
* invariant enforcement and frame generation (read/modifies).
*/

type T = int // for demo purposes 

// Sequence without duplicates
type  useq<T> =  s: seq<T> | !hasDuplicates(s)

// Checks if a sequence 's' has duplicates.
predicate hasDuplicates<T>(s: seq<T>)  {
    exists i, j :: 0 <= i < j < |s| && s[i] == s[j]
}

// Node of a binary search tree
class {:autocontracts} BSTNode {
    // Concrete implementation variables
    var value: T // value in this node
    var left: BSTNode?  // elements smaller than 'value'
    var right: BSTNode? // elements greater than 'value'
                        // (? - may be null)
    
    // Abstract variable used for specification & verification
    // purposes
    ghost var elems: set<T> // set of values in the subtree
            // starting in this node (including this value)

    // Class invariant with the integrity constraints for the
    // above variables
    predicate Valid() {  
      (elems == {value} 
                 + (if left == null then {} else left.elems)
                 + (if right== null then {} else right.elems))
      && (left != null ==> forall x :: x in left.elems 
                              ==> x < value)
      && (right != null ==> forall x :: x in right.elems 
                              ==> x > value)
      && (right != null ==> forall x :: x in right.elems 
                              ==> x > value)
      && (left != null ==> left.Valid())         
      && (right != null ==> right.Valid())
      && (left != null && right != null 
               ==> left.Repr !! right.Repr) 
        // disjoints sets of objects in left and right
        // subtree, needed to make sure that changing nodes 
        // in one subtree doesn't affect the other!    
    }

    // Initializes a new node with value 'x' and empty left 
    // and right subtrees.
    constructor (x: T)
      ensures elems == {x}
    {
        value := x;
        left := null;
        right := null;
        elems := {x};
    } 

    // Checks if the subtree starting in this node contains a
    // value 'x'. Runs in time O(log h), where 'h' is the 
    // height of the subtree.
    predicate method contains(x: T) 
      ensures contains(x) <==> x in elems
    {
      if x == value then true
      else if x < value && left!=null then left.contains(x)
      else if x > value && right!=null then right.contains(x)
        else false
    }

    // Inserts a value 'x' in the subtree starting in this 
    // node. If the value already exists, does nothing.
    // Runs in time O(log h), where h is the subtree height.
    method insert(x: T)
      ensures elems == old(elems) + {x}
      decreases elems
    {
        if  x == value {
            return;
        } 
        else if x < value {
            if left == null {
                left := new BSTNode(x);
            }
            else {
                left.insert(x);
            }
        }
        else {
            if right == null {
                right := new BSTNode(x);
            }
            else {
                right.insert(x);
            }
        }
        elems := elems + {x};
    }

    // Public function to find the maximum value in this
    // subtree. Runs in time O(log h), where 'h' is the
    // height of the subtree.
    function method max() : T
      ensures max() in elems 
              && forall x :: x in elems ==> x <= max()
    {
        if right == null then value else right.max()
    }

    // Public function to find the minimum value in this 
    // subtree.  Runs in time O(log h), where 'h' is the
    // height of the subtree.
    function method min() : T
      ensures min() in elems 
              && forall x :: x in elems ==> x >= min()
    {
        if left  == null then value else left.min()
    }

    // Deletes a value 'x' from the subtree starting in this
    // node, and returns the new head of the subtree (which
    // will be null if 'x' was the only value in the subtree).
    // If the value doesn't exist, does nothing.
    // Currently, seems to run in time O(log h * log h), where
    // 'h' is the height of the subtree.
    method delete(x: T) returns(res: BSTNode?)
      ensures if old(elems) == {x} then res == null 
              else res != null && res.elems == old(elems)-{x} 
                   && res.Valid() 
                        // not added  by autocontracts ...
                   && res.Repr <= old(Repr) 
                        // to preserve disjointness ...
      decreases elems
    {
        if  x == value {
           if left == null {
                res := right; // just changes the head
                return;
            }
            else if right == null {
                res := left; // just changes the head
                return;
            }
            else {
                if * /* non deterministic choice */ { 
                    value := left.max();
                    left := left.delete(value);
                }
                else {
                    value := right.min();
                    right := right.delete(value);
                }
            }
        } 
        else if x > value && right != null {
            right := right.delete(x);
        }
        else if x < value && left != null {
            left := left.delete(x);
        }
        res := this;
        elems := elems - {x};
    }

    method asSeq() returns (s: useq<T>)
      ensures isSorted(s) && asSet(s) == elems
      decreases elems
    {
        var l, m, r := [], [value], [];
        if left != null { l := left.asSeq(); }
        if right != null { r := right.asSeq(); }
        asSetProp(l, m); // recall lemma
        asSetProp(l + m,  r); // recall lemma
        return l + m + r;
    }
}

// Auxiliary function that obtains the set of elements in a
// sequence.
function asSet(s: seq<T>) : set<T> 
  ensures forall i :: 0 <= i < |s| ==> s[i] in asSet(s)
  ensures forall x :: x in asSet(s) ==> x in s
{ 
    if |s| == 0 then {} else {s[0]} + asSet(s[1..])  
}

// Auxiliary predicate that checks if a sequence is strictly
// sorted.
predicate isSorted(s: useq<T>) {
    forall i, j :: 0 <= i < j < |s| ==> s[i] < s[j]
}

// Lemma that states and proves by induction the following 
// property: the set of elements of sequence concatenation is
// the union of the individual sets of elements.
lemma asSetProp(s1: seq<T>, s2: seq<T>)
  ensures asSet(s1 + s2) == asSet(s1) + asSet(s2)
{
    if |s1| > 0 {
      assert s1 == s1[..1] + s1[1..];
      assert (s1[..1] + s1[1..]) + s2 == s1[..1]+(s1[1..]+s2);
      asSetProp(s1[1..], s2);
    } 
    else {
      assert [] + s2 == s2;
    }
}

// Lemma that states and proves by induction the following
// property: if two sequences without duplicates are sorted 
// and have the same set of elements, then they must be
// identical. 
lemma sortingUniqueness(a: useq<T>, b: useq<T>)
  requires isSorted(a) && isSorted(b) && asSet(a) == asSet(b)
  ensures a == b
{
    if |a| > 0 {
        sortingUniqueness(a[1..], b[1..]);
    }
}

// A simple test case.
method testSortedSet() {
    var s := new BSTNode(2);
    s.insert(5);
    s.insert(1);
    s.insert(4);
    s.insert(4);
    var t := s.asSeq();
    sortingUniqueness(t, [1, 2, 4, 5]); 
           // to help prove next assertion
    assert t == [1, 2, 4, 5];
    assert s.min() == 1;
    assert s.max() == 5;
    var s2 := s.delete(5);
    assert s2.elems == {1, 2, 4};
}
\end{lstlisting}

\subsection{Stable Marriage}
\small
\begin{lstlisting}
/* 
* Formal verification with Dafny of the Gale-Shapley algorithm
* to solve the stable marriage problem, both described in
* https://en.wikipedia.org/wiki/Stable_marriage_problem.
* Then, this algorithm is applied to solve the teachers
* placement problem that caused serious trouble in Portugal
* in 2004.
*/

// Sequence without duplicates
type  useq<T> =  s: seq<T> | !hasDuplicates(s)

// Injective map
type  inmap<K, V> =  m: map<K, V> | isInjective(m)

// Checks if a sequence 's' has duplicates.
predicate hasDuplicates<T>(s: seq<T>)  {
    exists i, j :: 0 <= i < j < |s| && s[i] == s[j]
}

// Checks if a map 'm' is injective, i.e., distinct keys are 
// mapped to distinct values. 
predicate isInjective<K,V>(m: map<K,V>) {
    forall i, j :: i in m && j in m && i != j ==> m[i] != m[j]
}

// Checks if element 'e1' precedes 'e2' in sequence 's'. 
predicate method precedes<T(==)>(e1: T, e2: T, s: seq<T>) {
  exists i, j :: 0 <= i < j < |s| && s[i] == e1 && s[j] == e2
}

// Obtains the set of elements in a sequence
function elems<T>(s: useq<T>): set<T> 
  ensures forall x :: x in elems(s) ==> x in s
  ensures forall x :: x in s ==> x in elems(s) 
{
    set i | 0 <= i < |s| :: s[i]
}

// Checks if a matching of couples is valid, i.e., men and 
// women can be engaged only if they are mentioned in each
// others preferences 
predicate isValid<Man, Woman>(couples: inmap <Man, Woman>,
            menPrefs: map<Man, useq<Woman>>, 
            womenPrefs: map <Woman, useq<Man>>) 
{
  forall m :: m in couples ==> var w := couples[m];
      m in menPrefs && w in womenPrefs 
      && w in menPrefs[m] && m in womenPrefs[w]
}

// Checks if a matching of couples is stable, i.e., there is
// no pair (m, w) that prefer each other as compared to their
// current situation. 
predicate isStable<Man, Woman>(couples: inmap <Man, Woman>,
            menPrefs: map<Man, 
            useq<Woman>>,
            womenPrefs: map <Woman, useq<Man>>) 
{
    ! exists m, w :: m in menPrefs.Keys 
          && w in womenPrefs.Keys
          && unstable(m, w, couples, menPrefs, womenPrefs)
}

predicate unstable<Man, Woman>(m: Man, w: Woman, 
            couples: inmap <Man, Woman>, 
            menPrefs: map<Man, useq<Woman>>, 
            womenPrefs: map <Woman, useq<Man>>)
  requires m in menPrefs.Keys && w in womenPrefs.Keys
{
    w in menPrefs[m] && m in womenPrefs[w] &&
    (m in couples ==> precedes(w, couples[m], menPrefs[m]))
    && (forall m' :: m' in couples && couples[m'] == w 
           ==> precedes(m, m', womenPrefs[w]))          
}

// Stable matching by the Gale-Shapley algorithm with
// incomplete lists and no ties.
// Receives the lists of preferences of men and women and
// returns the couples created.
// Time complexity (with proper data structures) is
// O(|M|*|W|), where W is the set of women and M the set of
// men. 
// The types Man and Woman are defined as type parameters
// because their internal structure is not relevant here.
method stableMatching<Man, Woman>(
           menPrefs: map<Man, useq<Woman>>, 
           womenPrefs: map <Woman, useq<Man>>) 
         returns(couples: inmap <Man, Woman>) 
  // P1: women referenced in men preferences must exist 
  requires forall m :: m in menPrefs 
        ==> forall w :: w in menPrefs[m] ==> w in womenPrefs
  // P2: man referenced in women preferences must exist 
  requires forall w :: w in womenPrefs 
         ==> forall m :: m in womenPrefs[w] ==> m in menPrefs
  // Q1: men and women can be engaged only if they are
  // mentioned in each other's preferences 
  ensures isValid(couples, menPrefs, womenPrefs)
  // Q2: stable marriage (and maximality)
  ensures isStable(couples, menPrefs, womenPrefs)
{
    // Initiate the result as empty
    couples  := map[];

    // Initialize the men preferences already explored empty
    var menPrefsExplored  := map m | m in menPrefs :: [];

    // Ghost variable used for proving termination with Dafny
    // (instead of menPrefsExplored, that has a too complex
    // structure)
    ghost var unexploredPairs := set m, w | m in menPrefs 
                               && w in menPrefs[m] :: (m, w);

    // while exists a free man m who still has a woman w to
    // propose to
    while exists m :: m in menPrefs && m !in couples 
                      && menPrefsExplored[m] < menPrefs[m]
      decreases unexploredPairs
      // I1: menPrefsExplored has the same keys (men) as
      // menPrefs 
      invariant menPrefs.Keys == menPrefsExplored.Keys
      // I2: lists in menPrefsExplored must be sublists
      // (prefixes) in menPrefs 
      invariant forall m :: m in menPrefsExplored 
                  ==> menPrefsExplored[m] <= menPrefs[m]
      // I3: to assure Q1 incrementally, with menPrefsExplored
      // instead of menPrefs 
      invariant isValid(couples, menPrefsExplored, womenPrefs)
      // I4: to assure Q2 incrementally, with menPrefsExplored
      // instead of menPrefs
      invariant isStable(couples,menPrefsExplored,womenPrefs)
      // I5: while engaged, men do not propose to further
      // women (needed to preserve isStable)
      invariant forall m :: m in couples 
                 ==> couples[m] == last(menPrefsExplored[m]) 
      // I6: inv. defining the contents of unexploredPairs
      invariant unexploredPairs == set m, w | m in menPrefs 
                        && w in menPrefs[m] 
                        && w !in menPrefsExplored[m] :: (m, w)
    {
        // select a man in such condition (free man m who
        // still has a woman w to propose to)
        var m :| m in menPrefs && m !in couples 
                 && menPrefsExplored[m] < menPrefs[m];  

        // select the next woman on m's list (using auxiliary
        // function to circumvent Dafny limitation)
        var w := nth(menPrefs[m], |menPrefsExplored[m]|); 

        // if w isn't free (i.e., some pair (m',w) exists yet)
        if m' :| m' in couples && couples[m'] == w 
        { 
            // if w prefers m to m'
            if m in womenPrefs[w] 
               && precedes(m, m', womenPrefs[w]) 
            {
                // m' becomes free
                couples := map x | x in couples 
                                   && x != m' :: couples[x];
                // (m, w) become engaged
                couples := couples[m := w];
            }
        }
        else // w is free
        {
            // if w is interested in m
            if m in womenPrefs[w]
            {
                // (m, w) become engaged
                couples := couples[m := w];
            }
        }        

        // mark this pair as explored
        menPrefsExplored := 
            menPrefsExplored[m := menPrefsExplored[m] + [w]];
        unexploredPairs := unexploredPairs - {(m, w)};
    }
}

/*
 * Some test cases for the stable marriage problem.
 */

method testStableMatching1() {
    var menPrefs := map [1 := [1, 2], 2 := [1, 2]]; 
    var womenPrefs := map [1 := [1], 2 := [2]];
    var expectedCouples := map[1 := 1, 2 := 2]; 
    var actualCouples := stableMatching(menPrefs, womenPrefs);
    assert isValid(expectedCouples, menPrefs, womenPrefs); 
             // proof helper...
    assert actualCouples == expectedCouples;
}

method testStableMatching2() {
    var menPrefs := map [1 := [2, 1], 2 := [1, 2]]; 
    var womenPrefs := map [1 := [1, 2], 2 := [2, 1]];
    var expectedCouples1 := map[1 := 2, 2 := 1]; 
    var expectedCouples2 := map[1 := 1, 2 := 2]; 
    var actualCouples := stableMatching(menPrefs, womenPrefs);
    assert isValid(expectedCouples1, menPrefs, womenPrefs); 
              // proof helper...
    assert actualCouples == expectedCouples1 
            || actualCouples == expectedCouples2;
}

method testStableMatching3() {
    var menPrefs := map [1 := [1, 2], 2 := [1]]; 
    var womenPrefs := map [1 := [1, 2], 2 := [2, 1]];
    var expectedCouples1 := map[1 := 2, 2 := 1]; 
    var expectedCouples2 := map[1 := 1]; 
    var actualCouples := stableMatching(menPrefs, womenPrefs);
    assert isValid(expectedCouples1, menPrefs, womenPrefs); 
             // proof helper...
    assert actualCouples == expectedCouples1 
            || actualCouples == expectedCouples2;
}

/* 
* Application to solve the teachers placement problem.
*/

type Teacher = nat
type Vacancy = nat

// Auxiliary function to move an element 'x' in a sequence 's'
// (without duplicates) to the head of the sequence.
function method moveToHead<T(==)>(s: useq<T>, x: T) : useq<T>
  requires x in s 
  ensures forall y :: y in s ==> y in moveToHead(s, x)
{
    var i :| 0 <= i < |s| && s[i] == x; [s[i]]+s[..i]+s[i+1..]
}

// Gets the last element in a sequence
function last<T>(s: seq<T>): T
requires |s| > 0
{ s[|s|-1] }

// Gets the n-th element in a sequence
function method nth<T>(s: seq<T>, n: nat): T
requires 0 <= n < |s|
{ s[n] }

// Auxiliary predicate that checks if a teacher 't' has
// precedence over the current teacher that occupies vacancy
// 'v', if any, knwowing the ranked list of teachers, their
// initial placement, and the final placement. 
// A teacher that initially occupied 'v' has priority over all
// others; otherwise, priority is given to teachers with
// higher rank. 
predicate method teacherHasPrecedenceForVacancy(t: Teacher, 
  v: Vacancy, finalPlacement: inmap<Teacher, Vacancy>,
  teachers: useq<Teacher>, 
  initialPlacement: inmap<Teacher,Vacancy>)
{
    if  t2 :| t2 in finalPlacement && finalPlacement[t2] == v 
    then t != t2 
        && ((t, v) in initialPlacement.Items 
           || ((t2, v) !in initialPlacement.Items 
                && precedes(t, t2, teachers)))
    else true // the vacancy is still free, so any teacher is
              // better than remaining free 
} 

// Solution for teachers placement problem, by reducing it to
// the stable marriage problem.
// Input parameters:
//   vacancies - set of vacancies available (includes
//               positions currently occupied by teachers that
//               want to change position)
//   teachers - ordered set of teachers, ordered by their
//              ranking (represented as a sequence without
//              duplicates)        
//   preferences - map that indicates for each teacher the
//                 ordered list of vacancies wanted
//   initialPlacement - map that indicates the initial
//                placement of teachers with initial placement
// Output parameters:
//   finalPlacement - final teachers placement
method teachersPlacement(vacancies: set<Vacancy>, 
          teachers: useq<Teacher>, 
          preferences: map<Teacher, useq<Vacancy>>,
          initialPlacement: inmap <Teacher, Vacancy>) 
       returns(finalPlacement: inmap<Teacher, Vacancy>) 
  // P1: the teachers in the ranked sequence and the teachers
  // with preferences, are the same
  requires elems(teachers) == preferences.Keys 
  // P2: the vacancies mentioned in teachers preferences must
  // exist in the set of vacancies
  requires forall t :: t in preferences 
              ==>  elems(preferences[t]) <= vacancies
  // P3: the teachers and vacancies mentioned in the initial
  // placement must exist in the sets of teachers and
  // vacancies 
  requires forall t :: t in initialPlacement 
        ==> t in teachers && initialPlacement[t] in vacancies
  // P4: the initial placement of a teacher must be mentioned
  // in his list of preferences as the last preference
  requires forall t :: t in initialPlacement ==> 
            initialPlacement[t] in preferences[t]
            && initialPlacement[t] == last(preferences[t])
  // Q1: the teachers mentioned in the final placement must
  // exist in the set of teachers
  ensures finalPlacement.Keys <= elems(teachers)
  // Q2: a teacher may only be placed in a vacancy mentioned
  // in his/her list of preferences 
  ensures forall t :: t in finalPlacement 
            ==> finalPlacement[t] in preferences[t]
  // Q3: the assignment is stable, i. e., there is no pair of
  // teacher t and vacancy v in his list of preferences,
  // such that t prefers v over his current situation (either
  // because t is free, or because t prefers v over the
  // assigned position), and v prefers t over its current
  // situation (either because v is free and so prefers any
  // teacher as compared to remaining free, or t is the
  // teacher initially placed and is not occupying v, or the
  // teacher t' that currently occupies v was not initially
  // placed there and has a lower rank than t) 
  ensures ! exists  t, v :: t in teachers 
             && v in preferences[t] 
             && (t in finalPlacement ==> precedes(v,
              finalPlacement[t], preferences[t]))//t prefers v
             && teacherHasPrecedenceForVacancy(t, v,
               finalPlacement, teachers, initialPlacement)
  // Q4: teachers that have an initial position must also have
  // a final position
  ensures forall t:: t in initialPlacement 
              ==> t in finalPlacement  
{
    // Reduction to the problem of stable marriage, with 
    // teachers as men (with the given preferences),
    // vacancies as women, and the preferences of each vacancy
    // given by the ranked list of teachers with the teacher 
    // initially placed there (if any) moved to the head   
    finalPlacement := stableMatching(preferences,
      vacanciesPrefs(vacancies, teachers, initialPlacement)); 
}

// preferences of each vacancy given by the ranked list of
// teachers with the teacher initially placed there (if any)
// moved to the head   
function method vacanciesPrefs(vacancies: set<Vacancy>,
           teachers: useq<Teacher>, 
           initialPlacement: inmap <Teacher, Vacancy>):
              map<Teacher, seq<Vacancy>>  
  requires forall t :: t in initialPlacement ==> t in teachers
                    && initialPlacement[t] in vacancies
{
    map v | v in vacancies ::
     if t :| t in initialPlacement && initialPlacement[t] == v 
     then moveToHead(teachers, t)
     else teachers 
}

/*
 * Some test cases for the teachers placement problem.
 */

method test1TP() {
  var vacancies := {1, 2};
  var teachers := [1, 2, 3];
  var preferences := map [1 := [2, 1], 2 := [1, 2], 3 := [2]];
  var initialPlacement :=  map [1 := 1];
  var expectedVacanciesPrefs := map[1 := [1,2,3], 
                                    2 := [1,2,3]];
  var expectedFinalPlacement := map[1 := 2, 2 := 1];
  var actualFinalPlacement := teachersPlacement(vacancies,
                   teachers, preferences, initialPlacement); 
  assert isValid(expectedFinalPlacement, preferences,
           expectedVacanciesPrefs); // proof helper...
  assert !unstable(1, 1, expectedFinalPlacement, preferences,
            expectedVacanciesPrefs); // proof helper...
  assert actualFinalPlacement == expectedFinalPlacement; 
}

method test2TP() {
  var vacancies := {1, 2};
  var teachers := [1, 2, 3];
  var preferences := map [1 := [2, 1], 2 := [1, 2], 
                          3 := [2, 1]];
  var initialPlacement :=  map [3 := 1];
  var expectedVacanciesPrefs := map[1 := [3, 1, 2], 
                                    2 := [1, 2, 3]];
  var expectedFinalPlacement := map[1 := 2, 3 := 1];
  assert moveToHead(teachers, 3) == [3, 1, 2]; // proof helper 
  var actualFinalPlacement := teachersPlacement(vacancies,
             teachers, preferences, initialPlacement); 
  assert isValid(expectedFinalPlacement, preferences,
            expectedVacanciesPrefs); // proof helper...
  assert !unstable(1, 1, expectedFinalPlacement, preferences,
            expectedVacanciesPrefs); // proof helper...
  assert actualFinalPlacement == expectedFinalPlacement; 
}
\end{lstlisting}

\subsection{Topological Sorting}
\small
\begin{lstlisting}
/*
 * Proof of correctness of the classic topological sorting
 * algorithm (Kahn's algorithm), simplified, in Dafny.
*/

// Defines a directed graph with vertices of any type T as a
// pair (V, E), where V is the vertex-set and E is the
// edge-set. 
// Each directed edge is represented by a pair of vertices. 
datatype Graph<T> = Graph(V: set<T>, E: set<(T,T)>) 

// Checks if G defines a valid graph (checks that E is a 
// subset of V*V).           
predicate validGraph<T>(G: Graph<T>) {
  forall e :: e in G.E ==> e.0 in G.V && e.1 in G.V
}

// Checks if a graph is acyclic.
predicate acyclic<T>(G: Graph<T>) {
   ! exists v :: v in G.V && existsSimplePath(G, v, v) 
}

// Check if there is a non-empty simple path from vertex u to
// vertex v in graph G. (Currently, 'simple' means without
// repeated edges, but could be without repeated vertices).
predicate existsSimplePath<T>(G: Graph<T>, u: T, v: T)
  decreases G.E
{
  (u, v) in G.E 
  || exists e :: e in G.E && e.0 == u 
       && existsSimplePath(Graph(G.V, G.E-{e}), e.1, v) 
}


// Removes a vertex v and its incident edges from a graph G.
function method removeVertex<T>(v: T, G: Graph<T>) : Graph<T> 
{
  Graph(G.V - {v}, set e | e in G.E && e.0 != v && e.1 != v)
}  

// Checks if a sequence s of vertices is a topological 
// ordering of the vertices of a graph G.
predicate isTopSorting<T>(s: seq<T>, G: Graph<T>) 
  requires validGraph(G)
{
  multiset(s) == multiset(G.V) 
  && forall i, j:: 0 <= i <= j < |s| ==> (s[j], s[i]) !in G.E  
}

// Checks if a vertex v in a graph G has incoming edges.
predicate method hasIncomingEdges<T>(G: Graph<T>, v: T) 
 requires v in G.V
{
  exists u :: u in G.V && (u, v) in G.E
}

// Topological sorting of the vertices of an acyclic directed
// graph. Returns a sequence (linear ordering) of the vertices
// in topological ordering.
method topsort<T>(G: Graph<T>) returns (s: seq<T>)
  requires validGraph(G) && acyclic(G)
  ensures isTopSorting(s, G)
{
  s := [];
  var R := G; // remaining graph
  while R.V != {} 
    // relation between s and R (basically, R = G - s)
    invariant R == Graph(set v | v in G.V && v !in s, 
                  set e | e in G.E && e.0 !in s && e.1 !in s)
    // s is a topological sorting of G - R 
    invariant multiset(s) == multiset(G.V - R.V) 
    invariant forall i, j:: 0 <= i <= j < |s| 
                 ==> (s[j], s[i]) !in G.E  
    // there are no edges from vertices in R to vertices in s 
    invariant forall i:: 0 <= i < |s| 
                ==> forall v :: v in R.V ==> (v, s[i]) !in G.E  
    decreases R.V
  {
    // recall property: a subgraph of an ayclic graph is also
    // acyclic 
    lemmaAcyclicSubgraph(R, G);

    // recall property: a vertex without incoming edges must
    // exist in a non-empty acyclic graph
    lemmaAcyclicIndegrees(R);

    // pick a vertex without incoming edges
    var v :| v in R.V && !hasIncomingEdges(R, v); 

    // append to the result
    s := s + [v];
     
    // remove that vertex and its outgoing edges from the 
    // graph
    R := removeVertex(v, R);
  }
}

/** SECOND LEMMA ***/

// States and proves by contradiction the following property: 
// a non-empty acyclic graph  must have at least one vertex
// without incoming edges (0 indegree).  
lemma lemmaAcyclicIndegrees<T>(G: Graph<T>) 
  requires validGraph(G) && G.V != {} && acyclic(G)
  ensures exists v :: v in G.V && !hasIncomingEdges(G, v)
{
  // For the sake of contradiction, assume that all vertices
  // have incoming edges.
  if forall v :: v in G.V ==> hasIncomingEdges(G, v) {
    // Then a path of any length can be built, possibly with
    // repeated edges and vertices, namely a path with length
    // |G.V| + 1 
    var p := genPath(G, |G.V| + 1);
    // Such a path must have repeated vertices, and
    // consequently at least one cycle
    lemmaPathLen(G, p);
    // So the graph is not acyclic, which contradicts the 
    // pre-condition
    assert ! acyclic(G);
  }
}

// Generates a valid path of a specified length n in a non
// empty graph G in which all vertices have incoming edges.
// Because of this property, a path with any length may be
// constructed (possibly with repeated edges and vertices). 
lemma genPath<T> (G: Graph<T>, n: nat) returns (p: seq<T>)
  requires validGraph(G) && G.V != {}  
  requires forall v :: v in G.V ==> hasIncomingEdges(G, v)
  ensures |p| == n && validPath(p, G)
{
  p := [];
  while |p| < n
    invariant |p| <= n && validPath(p, G)
  {
    var u :| u in G.V && (p == [] || (u, p[0]) in G.E);
    p := [u] + p;
  }
}

// Checks if a sequence p of vertices defines a valid  path
// (allowing repeated vertices and edges) in a graph G.
predicate method validPath<T>(p: seq<T>, G: Graph<T>) {
   forall i :: 0 <= i < |p| ==> p[i] in G.V
         && (i < |p| - 1 ==> (p[i], p[i+1]) in G.E)
}

// States and proves the property: given a graph G and a path 
// p in G, if the length of p exceeds the number of vertices
// then G has cycles.  
lemma lemmaPathLen<T>(G: Graph<T>, p: seq<T>)
  requires validGraph(G) && validPath(p, G) && |p| > |G.V| 
  ensures !acyclic(G)
{
  // first notice that, if all vertices are distinct, the 
  // length of the path cannot exceed the number of vertices
  // in G 
  if nodups(p) {
    lemmaSeqLen(p, G.V);
  }

  // consequently, there must exist repeated vertices in p, so
  // we pick a cyclic (complex) subpath 
  var i, j :| 0 <= i < j < |p| && p[i] == p[j];
  var p' := p[i .. j+1];

  // recall auxiliary lemma that assures that a simple cycle
  // also exist
  lemmaComplexPath(G, p');
}

// States and proves (by induction) the property: given any
// valid complex path p (possibly with repeated edges and/or
// vertices) in a graph G, there exists a simple path (without
// repeated edges) in G from the first to the last vertex in 
// the complex path. 
lemma lemmaComplexPath<T>(G: Graph<T>, p: seq<T>)
  requires G.V != {} && validPath(p, G) && |p| > 1
  ensures existsSimplePath(G, p[0], p[|p|-1])
  decreases p
{
  // handles case of first vertex repeated in the middle
  if i :| 1 <= i < |p|-1 && p[i] == p[0] {
    lemmaComplexPath(G, p[i..]);
  }
  // handles recursive case of proof by induction
  else if |p| > 2 {
    lemmaComplexPath(Graph(G.V, G.E - {(p[0],p[1])}), p[1..]);
  }
}

function elems<T>(s: seq<T>): set<T> {
  set x | x in s 
}

predicate nodups<T>(s: seq<T>) {
  forall i, j :: 0 <= i < j < |s| ==> s[i] != s[j]
}

// States and proves (by induction) the following property:
// the length of a sequence p of distinct elements from a set
// s cannot exceed the cardinality of the set.  
lemma lemmaSeqLen<T>(p: seq<T>, s: set<T>)
  requires nodups(p) && elems(p) <= s
  ensures |p| <= |s|
{
  if p != [] {
    lemmaSeqLen(p[1..], s - {p[0]});  
  }
}

/** FIRST LEMMA ***/

// States and proves (by contradiction) the following 
// property: a subgraph G of an acyclic graph G' is also
// acyclic.
lemma lemmaAcyclicSubgraph<T>(G: Graph<T>, G': Graph<T>) 
  requires validGraph(G) && validGraph(G') 
           && acyclic(G') && isSubGraph(G, G')  
  ensures acyclic(G)
{
  if ! acyclic(G) {
     var u :| u in G.V && existsSimplePath(G, u, u); 
       // exists, by the definition of acyclic
     lemmaExistsSimplePath(G, G', u, u); 
       // recall lemma implying that such a path also exists
       // in G
     assert !acyclic(G'); 
     // so G would not be acyclic, contradicting the precond.
  }
}

// Checks if a graph G is a subgraph of another graph G'.
predicate isSubGraph<T>(G: Graph<T>, G': Graph<T>) {
  G.E <= G'.E && G.V <= G'.V
}

// States and proves (by induction) the following property: if
// there is a (simple) path u-->v in a graph G and G is a
// subgraph of G', then a path u-->v also exists in G'.
lemma lemmaExistsSimplePath<T>(G: Graph<T>, G': Graph<T>, 
                               u: T, v: T)
  requires validGraph(G) && validGraph(G') 
           && isSubGraph(G, G') && existsSimplePath(G, u, v) 
  ensures existsSimplePath(G', u, v)
  decreases G.E
{ 
  if (u, v) !in G.E { // recursive case
    var e :| e in G.E && e.0 == u 
            && existsSimplePath(Graph(G.V, G.E-{e}), e.1, v); 
            // must exist by definition of existsPath
    lemmaExistsSimplePath(Graph(G.V, G.E-{e}), 
                          Graph(G'.V, G'.E-{e}), e.1, v); 
           // this lemma implies that 'e' also exist in G' 
  } 
}

/** Test cases ***/
method testTopSortingSingleSolution() {
  var G: Graph<nat> := Graph({1, 2, 3}, {(1, 2), (2, 3)});
  assert validGraph(G) && acyclic(G);
  var s : seq<nat> := [1, 2, 3];
  assert isTopSorting(s, G);  
  var t := topsort(G);
  assert t == s;
}

method testTopSortingMultipleSolutions() {
  var G: Graph<nat> := Graph({1, 2, 3}, {(1, 2), (1, 3)});
  assert validGraph(G) && acyclic(G);
  var s1 : seq<nat> := [1, 2, 3];
  var s2 : seq<nat> := [1, 3, 2];
  assert isTopSorting(s1, G);  
  assert isTopSorting(s2, G);  
  var t := topsort(G);
  assert t == s1 || t == s2;
}
\end{lstlisting}

\subsection{Eulerian Circuit}
\small
\begin{lstlisting}
/*
 * Proof of correctness of the Hierholzer algorithm (1873) to
 * find an Eulerian circuit in an Eulerian graph (method
 * findEulerCircuit).
 * Reference: https://en.wikipedia.org/wiki/Eulerian_path.
 */
 
/**** Graph representation and validity ****/
 
// Vertices can be of any type that supports equality.
type Vertex = nat // or other type
 
// Graph represented as a mapping from vertices to sets of
// adjacent vertices.
type Graph = m: map<Vertex, set<Vertex>> |
                  definesValidGraph(m)
 
// The mapping must be anti-reflexive and symmetric.
predicate definesValidGraph(m: map<Vertex, set<Vertex>>) {
  forall v, w :: v in m && w in m[v] 
         ==> w != v && w in m && v in m[w]
}
 
/**** Graph modification operations ****/
 
// Removes a vertex v from a graph G (if existent).
function rmvVertex(v: Vertex, G: Graph): Graph {
  map u | u in G && u != v :: G[u] - {v}
}
 
// Removes an edge (u, v) from a graph G (if existent).
function method rmvEdge(u: Vertex, v: Vertex, G: Graph): Graph 
  ensures var G' : Graph := rmvEdge(u, v, G);
          u in G && v in G && v in G[u] ==>
            hasEvenCard(G'[v]) != hasEvenCard(G[v]) 
            && hasEvenCard(G'[u]) != hasEvenCard(G[u]) 
{
  map x | x in G :: if x == u then G[x] - {v}
                    else if x == v then G[x] - {u} else G[x]
}
 
// Adds and edge (u, v) to a graph G.
function addEdge(u: Vertex, v: Vertex, G: Graph): Graph
  requires u in G && v in G && u != v
{
  map x | x in G :: if x == u then G[x] + {v} 
                    else if x == v then G[x] + {u} else G[x]
}
 
/**** Subgraphs ****/
 
// Check if G1 is a subgraph of G2 in terms of edges, but with
// the same vertex-set.
predicate isSubgraphE(G1: Graph, G2: Graph) {
  G1.Keys == G2.Keys && forall x :: x in G1 ==> G1[x] <= G2[x]
}
 
/**** Connectivity ****/
 
// Checks if a given graph is connected, i.e., there is a path
// between every two vertices.  
predicate isConnected(G: Graph) {
  forall u, v :: u in G && v in G 
     ==> connectedVertices(u, v, G)
}
 
// Checks if vertices u and v are connected in a graph G, 
// i.e., there is a path connecting them (without repeated
// vertices).
predicate connectedVertices(u: Vertex, v: Vertex, G: Graph)
  requires u in G && v in G
  decreases G
{
  u == v 
  || exists w :: w in G[u] 
       && connectedVertices(w, v, rmvVertex(u, G))  
}
 
// Proves by induction that if a vertex u belongs to a closed
// vertex-set C (under adjacency) in a graph G and v does not,
// then they must be disconnected.
lemma unconnectedVerticesLemma(u: Vertex, v: Vertex, G: Graph,
                               C: set<Vertex>)
  requires u in G && v in G && u in C && v !in C
  requires forall x :: x in C && x in G ==> G[x] <= C 
               // C is a closed vertex-set
  decreases G
  ensures !connectedVertices(u, v, G)
{
  // mimics the structure of connectedVertices
  forall w | w in G[u] {
    unconnectedVerticesLemma(w, v, rmvVertex(u, G), C);
  }  
}
 
/**** Vertex degrees ****/
 
// Checks if all vertices in a graph G have even degree (even
// number of incident edges).
predicate hasEvenDegrees(G: Graph) {
  forall v :: v in G ==> hasEvenCard(G[v])
}
 
// Checks if a set s has an even number of elements (even 
// cardinal).
predicate hasEvenCard<T>(s: set<T>) {
  |s| % 2 == 0
}
 
// If we remove from a graph G with even vertex degrees the
// edges of a subgraph T with even vertex degrees, we obtain a
// subgraph R with even vertex degrees.
lemma evenDegreesLemma(G: Graph, R: Graph, T: Graph)
   requires G.Keys == T.Keys == R.Keys
   requires forall x :: x in G 
             ==> T[x] <= G[x] && R[x] == G[x] - T[x]
   requires hasEvenDegrees(G) && hasEvenDegrees(T)
   ensures hasEvenDegrees(R)
{ // Thanks Dafny
}
 
/*** Walks, trails, circuits and augmentation properties ****/
 
// Checks if a sequence s of vertices defines a valid walk in
// a graph G.
predicate isValidWalk(s: seq<Vertex>, G: Graph)  {  
  (forall x :: x in s ==> x in G)
  && (forall i :: 0 <= i < |s| - 1 ==> s[i+1] in G[s[i]])
}
 
// Usefull augmentation property of valid walks.
lemma walkAugmentationLemma(s: seq<Vertex>, G: Graph, 
                            u: Vertex)
  requires isValidWalk(s, G) && u in G
           && (|s| == 0 || u in G[s[|s|-1]])
  ensures isValidWalk(s + [u], G)
{ /* Thanks Dafny */ }
 
// Checks if a sequence s of vertices traverses an edge 
// (u, v).
predicate traversesEdge(s: seq<Vertex>, u: Vertex, v: Vertex)
  requires u != v
{
  exists i :: 1 <= i < |s| && {s[i-1], s[i]} == {u, v}
}
 
// Useful augmentation property of traversed edges.  
lemma traversesEdgeProp(s: seq<Vertex>, v: Vertex)
  requires |s| > 0 && v != s[|s|-1]
  ensures traversesEdge(s + [v], s[|s|-1], v);
{
  // it seems the only thing Dafny needs is to show the 
  // "de-concatenation"
  assert var s' := s + [v]; s'[..|s|] == s && s'[|s|] == v;
}
 
// Checks if a sequence s of vertices defines a valid trail 
// in a graph G, i.e., a valid walk without repeated edges.
predicate isValidTrail(s: seq<Vertex>, G: Graph) {  
    isValidWalk(s, G)
    && forall i :: 1 <= i < |s| 
          ==> ! traversesEdge(s[..i], s[i-1], s[i])
}
 
// Usefull aumentation property of valid trails.
lemma trailAugmentationLemma(s: seq<Vertex>, G: Graph, 
                             u: Vertex)
  requires isValidTrail(s, G) && u in G
  requires |s| > 0 ==> u in G[s[|s|-1]] // to make valid walk
  requires |s| > 0 ==> !traversesEdge(s, s[|s|-1], u) 
             // to make valid trail
  ensures isValidTrail(s + [u], G)
{ /* Thanks Dafny */ }
 
 
// Checks if a sequence s of vertices defines a valid circuit 
// in a graph G, i.e., a non-empty trail in which the first
// and last vertices are identical.
predicate isValidCircuit(s: seq<Vertex>, G: Graph) {  
    isValidTrail(s, G) && |s| > 0 && s[|s|-1] == s[0]
}
 
// Shows that circuit augmentation (by embedding) implies the
// union of traversed edges.
lemma circuitAugmentationLemma1(s1: seq<Vertex>, i: int, 
         s2: seq<Vertex>, s3: seq<Vertex>, G: Graph)
  requires isValidCircuit(s1, G) && isValidCircuit(s2, G)
  requires 0 <= i < |s1| && s2[0] == s1[i] 
           && s3 == s1[..i] + s2 + s1[i+1..]
  ensures forall x, y :: x in G && y in G[x] ==>
      (traversesEdge(s3, x, y) <==> traversesEdge(s1, x, y) 
       || traversesEdge(s2, x, y))
{
    // it seems the only thing Dafny needs is to show the 
    // "de-concatenation" (slicing)
    assert s1[..i] == s3[ .. i];
    assert s2 == s3[i .. i + |s2|];
    assert s1[i + 1 ..] == s3[i + |s2|..];
}
 
// Proves by deduction that circuit augmentation, by embedding
// another circuit with disjoint edges, results in a valid
// circuit, without repeated edges.
lemma circuitAugmentationLemma2(s1: seq<Vertex>, i: int, 
         s2: seq<Vertex>, s3: seq<Vertex>, G: Graph)
  requires isValidCircuit(s1, G) && isValidCircuit(s2, G)
  requires 0 <= i < |s1| && s2[0] == s1[i] 
           && s3 == s1[..i] + s2 + s1[i+1..]
  requires forall k :: 1 <= k < |s1| 
             ==> !traversesEdge(s2, s1[k-1], s1[k])
  requires forall k :: 1 <= k < |s2| 
             ==> !traversesEdge(s1, s2[k-1], s2[k])
  ensures isValidCircuit(s3, G)
{
  // mimics the procedure for checking existence of duplicate
  // edges
  forall j, k | 1 <= j < k < |s3|  
    ensures {s3[k-1], s3[k]} != {s3[j-1], s3[j]}
    {
      // map to indices in original sequences
      var (j', sj) := if j <= i then (j, s1) 
                      else if j < i+|s2| then (j-i, s2) 
                      else (j-(|s2|-1), s1);
      var (k', sk) := if k <= i then (k, s1) 
                      else if k < i+|s2| then (k-i, s2) 
                      else (k-(|s2|-1), s1);
      // recall that edges are distinct in original sequences
      // (from pre-conditions)
      assert {sk[k'-1], sk[k']} != {sj[j'-1], sj[j']};
    }

  
}
 
/*** Euler trails and circuits ****/
 
// Checks if a sequence s of vertices defines an Euler circuit
// in a graph G, i.e., a circuit that traverses each edge of G
// exactly once.
predicate isEulerCircuit(s: seq<Vertex>, G: Graph) {
 isValidCircuit(s,G) // ensures no duplicate edge crossing
 && forall u,v :: u in G && v in G[u] 
          ==> traversesEdge(s, u, v)
}
 
// Proves by contradiction that a non-augmentable circuit r
// in a connected graph G must cover all edges, i.e., must be
// an Euler circuit.
lemma nonAugmentableCircuitLemma(G: Graph, r: seq<Vertex>)
  requires isConnected(G) && isValidCircuit(r, G)
  requires forall x, y :: x in r && y in G[x] 
              ==> traversesEdge(r, x, y)
  ensures isEulerCircuit(r, G)
{
   assert forall x, y :: x in r && y in G[x] ==> y in r; 
            // implied by 2nd precondition
   if v :| v in G && v !in r {  
     unconnectedVerticesLemma(r[0], v, G, set x | x in r);
     // this contradicts the hypothesis that G is connected, 
     // so v cannot exist; hence all vertices of G are covered
     // by r, and hence their incident edges (by 2nd pre).
  }
}
 
// Checks if a sequence s of vertices defines an Euler trail
// in a graph G, i.e., a trail that traverses each edge of G
// exactly once.
predicate isEulerTrail(s: seq<Vertex>, G: Graph) {
   |s| > 0 && isValidTrail(s, G)
   && forall x, y :: x in G && y in G[x] 
         ==> traversesEdge(s, x, y)
}
 
// Property of vertex degrees in an Euler trail: the number of
// incident edges on each vertex is even except for the first
// and last vertex if different.  
predicate EulerTrailDegrees(G: Graph, r: seq<Vertex>)
  requires isEulerTrail(r, G)
{
   var first, last := r[0], r[|r|-1];
   forall x :: x in G ==> 
     ((x==first) != (x==last) /*xor*/ <==> !hasEvenCard(G[x]))
}

// Proves by induction the above property about the vertex
// degrees in an Euler trail.  
lemma EulerTrailLemma(G: Graph, r: seq<Vertex>)
   requires isEulerTrail(r, G)
   ensures EulerTrailDegrees(G, r)
{
  if |r| > 1 {
    EulerTrailLemma(rmvEdge(r[0], r[1], G), r[1..]);
    }
}


/**** Main algorithms ****/

// Hierholzer algorithm to find an Euler circuit in a 
// non-empty Eulerian graph G based on depth-first search.
method findEulerCircuit(G: Graph) returns (r: seq<Vertex>)
  requires isConnected(G) && hasEvenDegrees(G) && |G| > 0
  ensures isEulerCircuit(r, G)
{
  // build initial circuit, starting in an arbitrary vertex,
  // and obtain remaining graph
  var v :| v in G;
  var R : Graph;
  r, R := dfs(v, G);
 
  // Ghost variable to help proving termination (vertices to
  // explore)
  ghost var V := set x| x in R && R[x] != {};
 
  // augment r as possible
  while exists i :: 0 <= i < |r| && R[r[i]] != {}
     // r is a valid circuit in G starting in v
     invariant isValidCircuit(r, G) && |r| > 0 && r[0] == v
 
     // R is a subgraph of G with even vertex degrees
     invariant hasEvenDegrees(R) && isSubgraphE(R, G)
 
     // R contains the edges not traversed by r in G
     invariant forall x,y::x in G && y in G[x] 
          ==> (y !in R[x] <==> traversesEdge(r,x,y))
 
     // V (variant) is the set of vertices that have adjacent
     // vertices not yet explored
     invariant V == set x | x in R && R[x] != {}
     decreases V
  {
     // select a vertex in r with outgoing edges to explore
     var i :| 0 <= i < |r| && R[r[i]] != {};
     var u := r[i];
 
     // memmorize old values needed later
     ghost var oldr, oldV := r, V;
 
     // do a DFS from this vertex in the remaining graph,
     // obtaining a new subcircuit and remaining graph
     var c : seq<Vertex>;
     c, R  := dfs(u, R);
 
     // insert the subcircuit in the main circuit
     r := r[.. i] + c + r[i + 1..];  
 
     // recall circuit augmentation properties to make sure
     // invariants are maintained
     circuitAugmentationLemma1(oldr, i, c, r, G); 
          // union of traversed edges
     circuitAugmentationLemma2(oldr, i, c, r, G); 
          // no duplicate edges
 
     // prove that the variant decreases
     V := set x | x in R && R[x] != {};
     assert u in oldV && u !in V;
     assert V < oldV;
  }
 
  // show that all edges of G have been traversed, because G
  // is connected
  nonAugmentableCircuitLemma(G, r);
}

// By performing a depth-first search, produces a complete
// valid trail in a graph G starting in a vertex v.
// Assuming all vertices in the graph have even degree, the
// produced trail is circular. 
// Returns the circular trail (r) and the
// remaining graph R (with unexplored edges). 
method dfs(v: Vertex, G: Graph) returns (r: seq<Vertex>, 
                                         R: Graph)
  requires hasEvenDegrees(G) && v in G  
  ensures isValidCircuit(r, G) && |r| > 0 && r[0] == v
  ensures isSubgraphE(R, G) && hasEvenDegrees(R)
  ensures forall x, y :: x in G && y in G[x] 
           ==> (y !in R[x] <==> traversesEdge(r, x, y))
  ensures R[v] == {} // all successors of v have been explored
{
    R := G; // subgraph with edges remaining to be visited
    ghost var T: Graph := map x | x in G :: {}; 
            // subgraph with  edges already traversed
    
    // Ghost variable to help proving termination (edges 
    // remaining)
    ghost var E := set x, y | x in G && y in G[x] :: (x, y);

    // initiate the result with the initial vertex (also last
    // vertex at this point) 
    r := [v];
    var u := v;

    // augment r as possible
    while R[u] != {}
      // R is a subgraph of G (with the same vertex-set)  
      invariant isSubgraphE(R, G) 
         
      // T is a subgraph of G with edges G - R  
      invariant T.Keys == G.Keys 
                && forall x :: x in G ==> T[x] == G[x] - R[x]

      // E is the set of edges in R 
      invariant forall x, y :: x in R && y in R[x] 
                                <==> (x, y) in E

      // r is a sequence of vertices starting in v and 
      // ending in u  
      invariant |r| > 0 && r[0] == v && r[|r|-1] == u

      // r travers exactly the edges in T (without repetions)    
      invariant isValidTrail(r, T) 
      invariant forall x, y :: x in G && y in G[x] 
                ==> (y in T[x] <==> traversesEdge(r, x, y))
 
      // variant to ensure termination
      decreases E
    {
        // select an adjacent vertex following an edge not 
        // previously visited 
        var w :| w in R[u];

        // recall some trail augmentation properties  
        trailAugmentationLemma(r, G, w); // valid trail
        traversesEdgeProp(r, w); // traversed edges

        // augment the trail and update visited and 
        // non-visited edges and last vertex
        r := r + [w];
        R := rmvEdge(u, w, R);
        T := addEdge(u, w, T);
        E := E - {(u, w), (w, u)};
        u := w;

    }

    // shows that the obtained trail (Euler trail in T) ends
    // in the start vertex 
    assert T[u] == G[u]; // because R[u] == {}
    assert hasEvenCard(T[u]); // because hasEvenCard(G[u])
    EulerTrailLemma(T, r);
    assert u == v;

    // shows that in the remaining graph (R) all vertices have
    // even degrees 
    assert hasEvenDegrees(T);
    evenDegreesLemma(G, R, T);
    assert hasEvenDegrees(R);   
}

method testEulerCircuit() {
  var G : Graph := map[1 := {2, 3}, 2 := {1, 3}, 
                 3 := {1, 2, 4, 5}, 4 := {3, 5}, 5 := {3, 4}];
  var c : seq<Vertex> := [1, 2, 3, 4, 5, 3, 1];   
  assert c == [c[0], c[1], c[2], c[3], c[4], c[5], c[6]]; 
                  // helper ...
  assert isEulerCircuit(c, G);
}

method testEulerTrail() {
  var G : Graph := map[1 := {2, 3}, 2 := {1, 3}, 
            3 := {1, 2, 4}, 4 := {3, 5}, 5 := {4}];
  var c : seq<Vertex> := [3, 2, 1, 3, 4, 5];   
  assert c == [c[0], c[1], c[2], c[3], c[4], c[5]]; 
                   // helper ...
  assert isEulerTrail(c, G);
}
\end{lstlisting}

\end{document}